\documentclass[%
 reprint,
nofootinbib,
 amsmath,amssymb,
 aps,
 pre,
]{revtex4-2}

\usepackage{graphicx}
\usepackage{dcolumn}
\usepackage{bm}
\usepackage{hyperref}

\usepackage[font=small,labelsep=period, justification=raggedleft, singlelinecheck=false]{caption}
\usepackage{wrapfig}
\usepackage{ragged2e}

\usepackage{diagbox}
\usepackage{multirow}

\begin{document}

\preprint{APS/123-QED}

\title{Bananas for Tonks' gas: structural transitions in bent-core molecule bi-layers}

\author{David A. King}
\email{daking2@lbl.gov}
\affiliation{Department of Physics and Astronomy, University of Pennsylvania, 209 South 33rd St., Philadelphia, PA, 19104, USA
}
\altaffiliation[Currently at: ]{Department of Materials Science and Engineering, University of California, Berkeley, CA 94720, USA
}
\altaffiliation{Materials Science Division, Lawrence Berkeley National Laboratory, Berkeley CA 94720, USA
}
\date{\today}

\begin{abstract}
We study the role of particle geometry in the ordering of bilayers of bent-core particles using a simplified two-dimensional model. Particles are confined to two parallel one-dimensional layers and can adopt only two discrete orientations of their bent cores. By mapping the system exactly onto a pair of coupled Ising chains in the thermodynamic limit, we may obtain closed-form expressions for all correlation functions. The coupling between orientational and layer degrees of freedom produces remarkably rich behaviour, including crossovers between ferromagnetic and antiferromagnetic orientational order, and states with polarised layers. Despite its simplicity, this exactly soluble model reproduces qualitative features of bent-core smectic liquid crystals and confined colloids, providing insight into the subtle role of particle geometry in driving structural correlations. 
\end{abstract}

\maketitle
\section{Introduction}
Compromise is a difficult art to master yet, despite their simplicity, hard-core particles come close. Any individual particle values its personal space, preferentially picking positions that provide plentiful possibilities in order to maximise entropy. When many such particles are brought together, they must compromise their individual preferences for the greater good; if one particle is too greedy for space, its neighbours will lose out, and the entropy of the whole system decreases. This act of entropic compromise, repeated many thousands of times over, can result in a dizzying array of phases and micro-structures \cite{Frenkel2015OrderEntropy, Kamien2014}. 

Naturally, these structures depend on the geometry of the particles considered. Remarkably, even maximally symmetric hard-spheres have a non-trivial phase diagram, possessing both fluid-like and crystalline structures \cite{Pusey2009HardFormation,Wood1957PreliminarySpheres,Alder1957PhaseSystem}. More complicated structures can arise when the symmetry of the particles is reduced, and liquid crystalline phases can emerge between the solid and fluid \cite{DeGennes1993a,Frenkel1991}. Notably, as their concentration increases but before they crystallise, \textit{any} collection of sufficiently anisotropic particles will form a nematic liquid crystal \cite{Onsager1949,Kamien2014}. The compromise made here is between translational and orientational entropy; to keep their positions homogeneously distributed and maximise their translational freedom, the particles sacrifice their orientational entropy and align their long axes with their neighbours. 

If the particle density is increased still further, another liquid crystal phase can appear, the smectic-A phase. Having already sacrificed their orientations, the particles have only their freedom in different directions with which to bargain. This results in them organising in layers perpendicular to their orientations; the particles gain space within their layers and keep their positions random in this direction by trading in freedom to move perpendicular to the layer. The smectic phase is more elusive than the nematic, only being coaxed out by certain particle geometries \cite{Frenkel1984,Stroobants1986,Bolhuis1997TracingSpherocylinders,Evans1992}. For the smectic-A phase, the most important parts of the particles are the ``tips'' lying \textit{between} the layers \cite{ Cacelli2007,King2023WhatEnds,King2024MyDimensions}. 

Once the smectic has formed, the geometry of the particles \textit{within} the layer becomes important. In particular, if the particles are bent like bananas then, as the particles negotiate among themselves their freedom to point their elbows as they please, a range of structural possibilities emerge \cite{Jakli2018PhysicsMolecules}. Most notably, the layers can become ``polarised'', with their constituent particles having a bias to point in a particular direction. This raises a new issue for the particles in a given layer to debate; ``should we point our polarisation in the same way as our neighbours'?'' If the ayes have it, all layers assume \textit{the same polarisation}. This phase is referred to as ``$\text{SmA-P}_{\text{F}}$'' \cite{Niori1996DistinctMolecules,Sekine1997FerroelectricSystems,Guo2011FerroelectricMolecules,Reddy2011SpontaneousLayers}. The first three letters of the name indicate that the phase is a kind of smectic-A phase, the third, $\text{P}$, indicates that the layers have polar order for the particles directions, and the subscript indicates how this polar order compares between subsequent layers: $\text{F}$ for ``ferromagnetic'' by analogy to the ordering of spins in a magnet \footnote{``Ferroelectic'' is more commonly used for this phase. We shall map our system onto a spin chain, where the magnetic analogy is more natural.}. The particles could also choose contrarianism, resulting in all layers having \textit{opposite} polarisations to their neighbours. Repeating the spin analogy, this is the $\text{SmA-P}_{\text{A}}$ phase, with $\text{A}$ standing for ``antiferromagnetic'' \cite{Eremin2001ExperimentalBehavior, Guo2011TransitionCrystals}. 

Each of these phases has been observed experimentally and have their own suite of curious electro-optic responses (for a detailed review see \cite{Jakli2018PhysicsMolecules}). On top of characterising the phases, it is important to understand which particles form what phases and why. Unfortunately, this is easier said than done and, in general, the understanding is qualitative and heuristic. For instance, the fact that layers are generally naysayers and $\text{SmA-P}_{\text{A}}$ is more common than $\text{SmA-P}_{\text{F}}$, can be understood intuitively based on the restrictions to out of layer fluctuations imposed by the anticlinic particle orientations between layers \cite{Jakli2018PhysicsMolecules,Reddy2011SpontaneousLayers}. However, it is difficult to understand this and its relation to the particle shapes in detail without appealing to phenomenological models.

In this paper, we aim to understand the relationship between particle geometry and phase behaviour in a simplified model system of bent-core particles which may be solved \textit{exactly}. In order for the model to be tractable, the details of true smectics must be stripped back to their bare necessities. We consider a system of identically shaped, bent-core particles in two dimensions that are restricted to lie on a \textit{pair} of parallel, one-dimensional layers. The particles interact only by excluded volume and we restrict each to one of two orientations: left or right. In the thermodynamic limit, we show that this system is equivalent to a one-dimensional spin chain, where each site corresponds to a particle and carries two Ising variables: one indicates the orientation and the other the layer. These kinds of spin models have been previously been proposed phenomenologically to model structures of smectic phases \cite{Selinger2003ChiralCrystals,Reddy2011SpontaneousLayers}. Here, however we are able to derive the model precisely allowing for a closer connection between the phase behaviour and particle geometry. 

While our model may be solved exactly, it is so pared-down that any application to true smectics is still inherently qualitative. Nevertheless, it is still valuable as a solid foundation upon which to rest intuitive arguments and build more detailed realistic models. Furthermore, simplified models of smectics as particles distributed between a limited number of one-dimensional layers have been able to capture the important relationship between particle tip shape and smectic stability \cite{King2023WhatEnds,King2024MyDimensions}. Similarly, arguably the most important interactions for determining the structure of bent-core smectics are those between particles on neighbouring layers. By restricting our attention to two layers, we are able to probe these interactions in a setting where detailed analysis is still possible. 

Beyond the potential application to smectic liquid crystals, the model is highly relevant to systems of strongly confined particles, where uncompromising walls disrupt the cordial entropic debate among the particles \cite{Lowen2009TwentyBreathing}. This leads to a rich array of structures not seen in bulk systems \cite{Fortini2006PhaseConfinement}. Particularly notable are the quasi-one-dimensional structures produced in cylindrical confinement, where novel structures of water ice and helical packings of hard spheres are found \cite{Lohr2010HelicalCylinders,Koga2001FormationNanotubes,Fu2016HardCylinders}. Particle geometry is, of course, important for these structures too: hard spheres form zig-zags under these conditions while hard rods organise into two parallel rows as pressure increases \cite{Gurin2018PositionalNanopores}. Interpreting these rows as our pair of layers, the model presented here fleshes out this behaviour for bent-core particle bilayers and our exact solution shows it to be surprisingly rich. 

Despite the one-dimensional character of the system ruling out any true phase transitions \cite{vanHove1950SurDimension}, the correlation functions can have non-trivial behaviour. In particular, we find that, depending on the particle geometry, the correlations between the orientations of different particles can swap from ferromagnetic to antiferromagnetic. These correlations always decay as a single exponential with the separation between the particles considered. On the other hand, the correlations between the particle layers (and the layer-orientation cross-correlations) have a detailed multi-exponential structure and show zig-zag-like or layer-segregating behaviour. This reflects the subtle interplay between orientational and layering order in this model system, highly confined systems or true smectic liquid crystals.  

We begin by describing the model in detail in section \ref{sec:System}. Here we introduce and derive the central quantity in this work, the ``structural probability distribution'' which describes how the particles' orientations and how they are distributed between the layers. We show that, in the thermodynamic limit, this becomes a Boltzmann distribution, with an effective temperature set by the packing fraction of particles and an effective Hamiltonian equivalent to that of two coupled Ising models. This structure allows us to compute expectation values and correlation functions \textit{exactly}. 

In section \ref{sec:OrCorr} we obtain an exact expression for the correlation between the orientations of distant particles. In section \ref{sec:Bans}, we present a single parameter family of banana shaped particles which exhibit all possible behaviours of the orientational correlations. In section \ref{sec:LLCorr} we compute the correlations between the layers occupied by distant particles, as well as the cross-correlation between the layers and orientations. The behaviours of these correlations are significantly more complicated than the orientational correlations and in section \ref{sec:LLCorrBan} we map them out for the family of bananas introduced in section~\ref{sec:Bans}. The exact calculations presented throughout the paper are quite involved and do not present much physical insight beyond their results. For this reason we present a more intuitive discussion in section~\ref{sec:PhysDisc}, where we aim to understand the ordering shown by the correlations physically before we conclude in section~\ref{sec:conc}.
\section{Description of System and Formulation}
\label{sec:System}
\begin{figure*}  
\includegraphics[width=\textwidth]{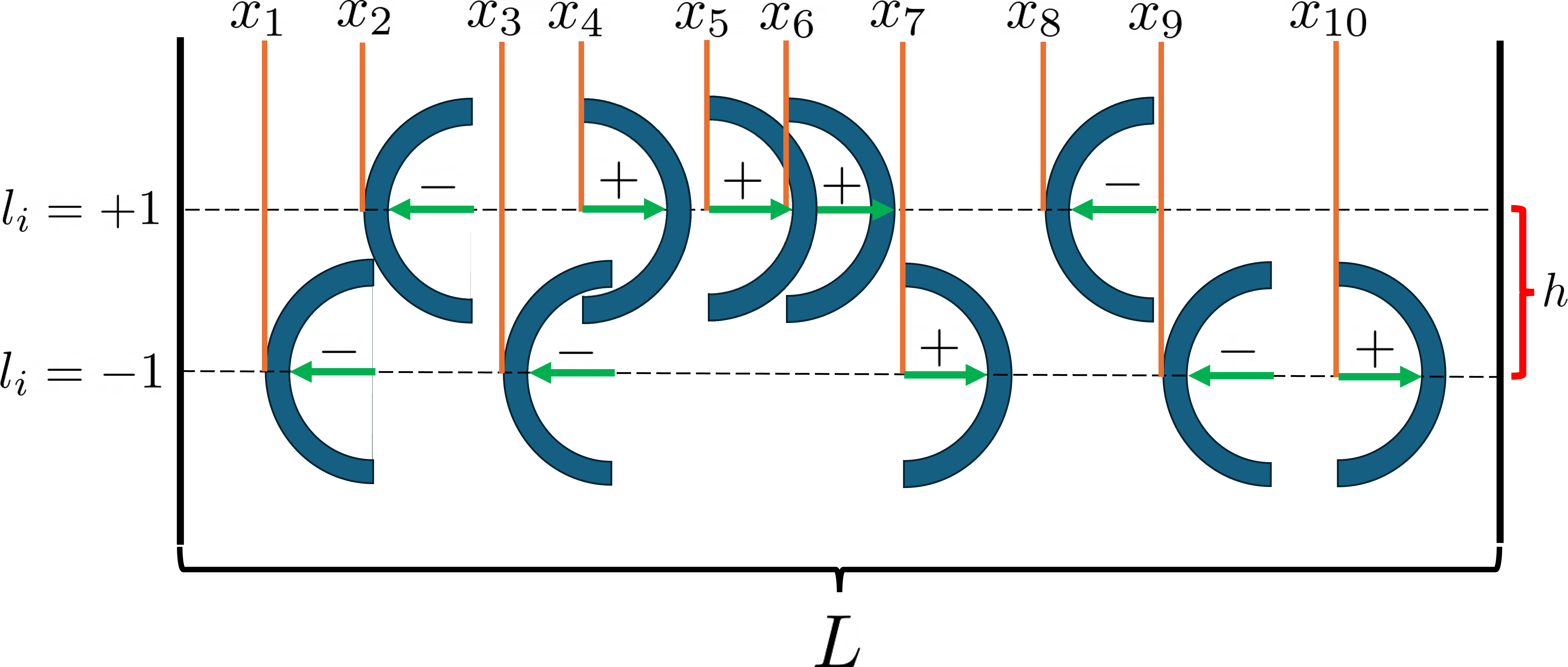}
\caption{\justifying A sketch of the kind of system we consider. A total of $N$ particles, shown in blue, are distributed between the two dashed layers which are separated by a height $h$, shown in red, and have total length $L$ with hard walls at their ends. Here we show a small number, $N=10$, particles but we consider systems in the thermodynamic limit: $N \to \infty$ and $L \to \infty$ with $N/L = \rho$ fixed. The layer on which the $i^{th}$ particle sits is labelled buy $l_i$; for the lower layer, $l_i=-1$, while for the upper $l_i = + 1$. The horizontal position of the \textit{left-most} point on the $i^{th}$ is given by $x_i$ and is marked in orange for each particle. The particles are not left-right symmetric, and can be in one of two orientations given by $\sigma_i = \pm 1$ indicated by the green arrows. Our convention puts $\sigma_i= +1 $ if the arrow points right and $\sigma_{i}=-1$ if it points left. }
\label{fig:system}
\end{figure*}

\begin{figure}\includegraphics[width=8cm]{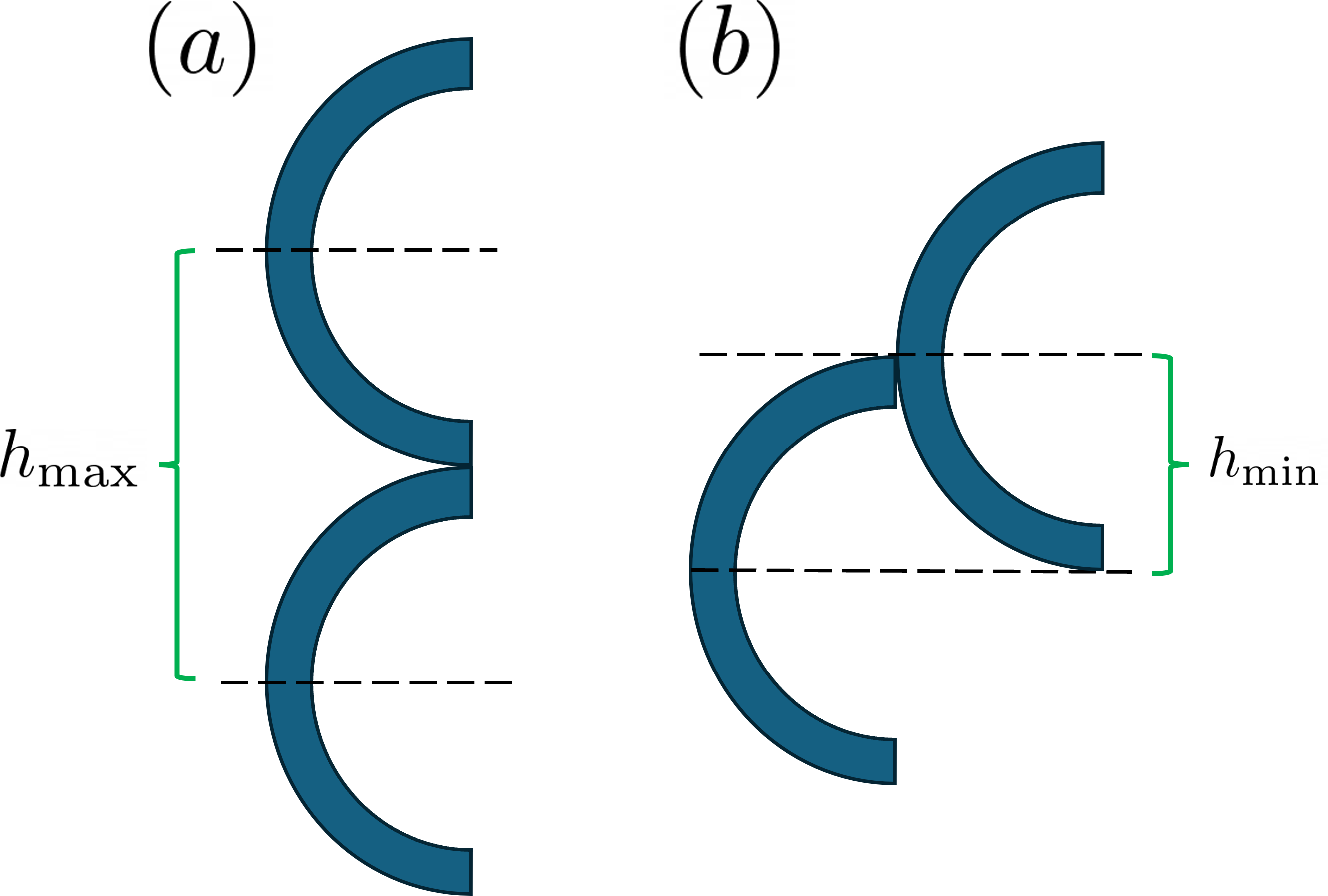}
\captionof{figure}{\justifying \label{fig:hmaxmin} Sketches showing the (a) maximum and (b) minimum layer spacings we consider. The maximum layer spacing, $h_{\text{max}}$, is set to that particles on different layers only just touch. The minimum, $h_{\text{min}}$, is set so that the lowest point of the upper particle is level with the centreline of lower particle. }
\end{figure}

We consider the system as shown in Fig.~\ref{fig:system}, where $N$ identical particles are distributed between two horizontal layers of equal length, $L$. The layers are one-dimensional, parallel and separated by a distance $h$. The particles interact only by excluded volume, allowing them to occupy any horizontal position with equal probability, so long as they do not overlap any other particle or the walls at $x=0$ and $x=L$. When a particle sits on a layer, it is perfectly bisected by that layer. We only consider the case where the layers are sufficiently close that two particles on different layers can touch in the interstice, but far enough apart that the bottom-most point of the upper particle does not fall below the level of the lower layer. If the total vertical size of the particle is $\mathcal{L}$, then the layer spacing falls in the range: $\mathcal{L}/2={h_{\text{min}}\leq h\leq h_{\text{max}}=\mathcal{L}}$, as shown in Fig.~\ref{fig:hmaxmin}.

For simplicity, we shall only consider certain particle shapes: They must be up-down symmetric, i.e. they have mirror symmetry about the layer on which they sit, but are not necessarily left-right symmetric. Combined with the two-dimensional nature of the system, these constraints limit each particle to two distinct orientations. In either orientation, let $D_{\text{LR}}$  denote the horizontal distance from the \textit{absolute} left-most to the right-most point on the particle at a vertical displacement $z$ from the central mirror plane. By symmetry, $D_{LR}$ must be \textit{even} in $z$. We limit ourselves to particles for which $D_{\text{LR}}(z)$ is \textit{monotonic}, either increasing or decreasing, for $z>0$. Symmetry demands $z=0$ is an extremum of $D_{\text{LR}}(z)$ and we shall refer to this point as the particle's ``elbow''. An illustrative example of this kind of particle is the ``C-shape'' shown in Fig.~\ref{fig:system}.

We would like to understand both the distribution of the particles between the two layers and their orientations. Since each particle is in one of two orientations, left or right, the orientation of the $i^{th}$ particle can be described by the Ising-like variable $\sigma_{i}=\pm 1$. We take $\sigma_{i} = +1$ if its elbow points right, and $\sigma_{i} = - 1$ if it points left. We may also introduce another such variable, $l_i$, to indicate which layer the particle occupies; $l_i = 1$ if it is on the upper layer and $l_{i} = -1$ for the lower layer. We shall compile the orientations of the particles into the vector $\textbf{S} = (\sigma_1, \cdots , \sigma_{N})$ and their layers into the vector $\textbf{L} = (l_1 , \cdots, l_N)$. Here we have taken the particle labelled ``$1$'' as the left-most particle and ``$N$'' as the right-most. We can also compile the horizontal positions of the particles on their layers into the vector $\textbf{X} = (x_1, \cdots, x_N)$. For convenience, the position $x_i$ is defined as the position of the \textit{left-most} point on $i^{th}$ particle, not that of its geometric centre. Every value of $x_i$ which does not lead to two particles overlapping is assumed to be accessed with equal probability. Our goal is to determine the probability of a particular set of orientations and layer occupancies for the particles, $P(\textbf{S}, \textbf{L})$, independent of their positions. This describes the detailed structure of the system, and will allow us to assess the ordering of the particles orientations within and between the layers.

\subsection{Structural Probability Distribution}
\label{sec:PF}
To obtain $P(\textbf{S},\textbf{L})$, we start from Gibbs' definition of the free energy, $F = -k_B T \sum_{\textbf{S},\textbf{L}}\int d\textbf{X} \mathcal{P} \log \mathcal{P}$, where $\mathcal{P}$ is the probability of realising a particular set of orientations, layers and positions. Then we use the definition ${\mathcal{P} = P(\textbf{S},\textbf{L}) P_c(\textbf{X} | \textbf{S}, \textbf{L})}$, which introduces the conditional probability, $P_c$, of a set of positions \textit{given} a particular set of orientations and layer occupancies. The form of this conditional probability follows from our assumption that each particle accesses every allowed position with equal probability: $P_c = \Theta(\textbf{X}|\textbf{S}, \textbf{L})/ Z(\textbf{S}, \textbf{L})$. Here $\Theta$ is a unit indicator function picking out all allowed positions of the particles for a given set of orientations $\textbf{S}$ and layers $\textbf{L}$, and the function $Z(\textbf{S},\textbf{L}) = \int d\textbf{X} \Theta(\textbf{X}|\textbf{S}, \textbf{L})$ provides the nomalisation. We determine the equilibrium distribution $P(\textbf{S},\textbf{L})$ by minimising the free energy, which results in 
     \begin{equation}
     \label{eq:PZ}
         P(\textbf{S},\textbf{L}) \propto Z(\textbf{S},\textbf{L}),
     \end{equation}
where the constant of proportionality is determined by nomalisation. This shows that $Z$ controls the orientational order and layer occupancy of the system. This function is the configurational integral for $N$ non-overlapping particles distributed between the two layers according to $\textbf{L}$ with the orientations set by $\textbf{S}$. As we are able to uniquely order the particles from left to right, this system is effectively one-dimensional. Furthermore, the particles only interact by excluded volume. Hence $Z$ is the partition function of an ``embellished'' Tonks gas \cite{Tonks1936}. Tonks studied a one dimensional gas of particles which interact only by their excluded volume, our embellishment of this system takes into account the different orientations of neighbouring particles as well as which layers they occupy.

The crucial quantity in Tonks', and our, calculation is the length excluded between neighbouring pairs of particles. This will, of course, depend on their orientations and upon which layers they each sit. We shall define $\mathcal{W}(i , i+1)$ as smallest separation between the leftmost points on the $i^{th}$ and $(i+1)^{st}$ particles as particle $i$ approaches from the left. For notational clarity, its dependence on $\sigma_i$ and $l_i$ has been kept implicit. The particular form of $\mathcal{W}$ depends on the geometry of the particles, but its absolute size is set by the total width of the particles from left to right at their widest, $w$. Going forward, we consider the non-dimensional ${W(i,i+1) = \mathcal{W}(i,i+1)/w}$, interpreted as the excluded length measured in units of the particle width.

The unique ordering of the particles from left to right lets us follow Tonks' original calculation precisely and, \textit{mutatis mutandis}, find an equivalent result. The form of the partition function is \textit{the same} as that for a one-dimensional ideal gas. However, instead of being allowed to visit the entire length $L$, the excluded volume interactions between the particles effectively allow each one to only access a ``free length'' equal to the total length minus the sum of all the mutual excluded lengths \cite{Tonks1936,Kamien2014}. This sum includes the length excluded by the walls at $x=0$ and $x = L$. Hence, we have
\begin{equation}
\label{eq:PFGen}
    Z(\textbf{S},\textbf{L}) = \frac{1}{N!} \left(L - w \sum_{i=0}^{N} W(i,i+1)\right)^{N}.
\end{equation}

To proceed, we require an expression for the reduced excluded lengths $W(i,i+1)$. This may be constructed for any particle satisfying the requirements discussed previously by considering symmetry alone, as we shall later. For now, we note that there generally exists a ``maximally dense packing'' where $N_{\text{max}}$ particles, with orientations $\textbf{S}^*$ on layers $\textbf{L}^*$, use up all the available length allowing no more to be added\footnote{There will be \textit{multiple} such configurations. For instance, all the orientations could be reversed or the layers swapped. This does not affect our argument.}. 

It is difficult to determine this maximally dense packing in general. Therefore, we would like to re-write (\ref{eq:PFGen}) in such a way that we \textit{do not} require knowledge of its details. To this end, let us define ${N_{\text{max}} W_{\text{max}}= \sum_{i=0}^{N_{\text{max}}} W^*(i,i+1)}$, where ${W^*(i,i+1)}$ is the excluded length between neighbouring particles in the maximally dense packing and $W_{\text{max}}$ measures the ``excluded length per particle at maximum packing''. If we define the ``packing fraction'' of particles as ${\varphi = N w/L}$, which attains a maximum value of ${\varphi_{\text{max}}= N_{\text{max}} w/L}$, then we can express $W_{\text{max}}$ in terms of $\varphi_{\text{max}}$. 

By definition, no more particles can be added in the maximally dense packing. Therefore, the total excluded length between the particles must be the entire system length, save an amount smaller than a single particle, i.e. $w{N_{\text{max}} W_{\text{max}} = L- b w}$ where ${b < 1}$.  It follows that ${W_{\text{max}} = \varphi_{\text{max}}^{-1} (1 - b w/L)}$. In the thermodynamic limit, (${N,L \to \infty}$ with fixed packing fraction ${\varphi}$) each individual particle becomes vanishingly small compared to the size of the system; $w/L \to 0$. Hence, the excluded length per particle at maximum packing is simply the inverse of the maximum packing fraction: $W_{\text{max}} \sim \varphi_{\text{max}}^{-1}$. 


With this in hand, we can now simplify the configurational integral~\eqref{eq:PFGen} in the thermodynamic limit. First, add and subtract $w N W_{\text{max}}$ inside the bracket in (\ref{eq:PFGen}), resulting in 
\begin{equation}
        Z(\textbf{S},\textbf{L}) = \frac{1}{N!} \left(L - w N W_{\text{max}} - w\mathcal{H}(\textbf{S},\textbf{L})/4\right)^{N},
\end{equation}
where, introducing a convenient factor of four, we have defined ${\mathcal{H}(\textbf{S},\textbf{L}) = 4 \sum_{i} [W(i, i+1) - W_{\text{max}}]}$. By pulling a factor of ${L - w N W_{\text{max}}}$ out from the bracket above, we see that it becomes an exponential in the thermodynamic limit
\begin{equation}
        Z(\textbf{S},\textbf{L}) \to \frac{L^N}{N!} (1 - \varphi/\varphi_{\text{max}})^{N} \exp\left(- \beta \mathcal{H}(\textbf{S},\textbf{L})\right).
\end{equation}
From (\ref{eq:PZ}) the probability $P(\textbf{S},\textbf{L})$ becomes the Boltzmann distribution 
\begin{equation}
\label{eq:PBolt}
   P(\textbf{S},\textbf{L}) = \frac{1}{\mathcal{Z}}\exp\left(-\beta \mathcal{H}(\textbf{S},\textbf{L})\right),
\end{equation}
with the effective inverse temperature ${\beta = \varphi /(4 -4\varphi/\varphi_{\text{max}}) \neq (k_B T)^{-1}}$, effective Hamiltonian $\mathcal{H}$ and partition function $\mathcal{Z}$ required for normalisation. 

All expectation values derived from this distribution will be functions of $\beta$, which is small at low packing fractions and diverges when the packing fraction becomes maximal. This encodes the density of the system, but crucially we do not need to know the maximally dense packing explicitly. Note that \textit{both} the effective temperature and Hamiltonian depend on the particle geometry under consideration. Furthermore, any constant terms in the Hamiltonian will not affect expectation values of functions of $\textbf{S}$ or $\textbf{L}$, and shall be dropped.    

\subsection{Lengths Excluded \& Hamiltonian Affected}
\label{sec:LengExc}

\begin{figure}\includegraphics[width=8cm]{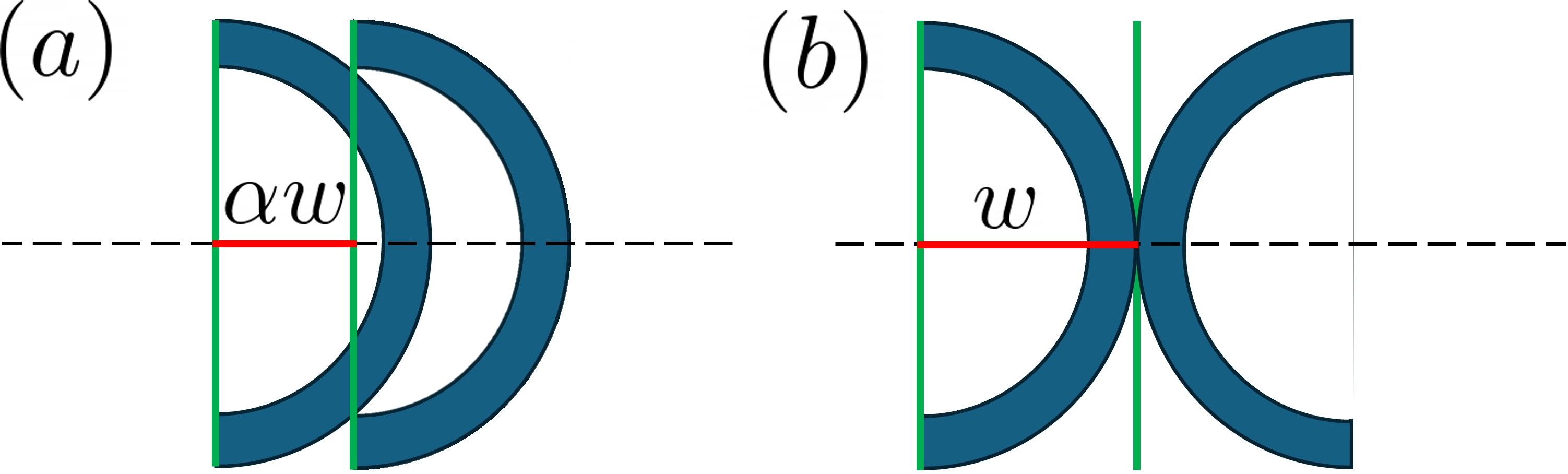}
		\captionof{figure}{\justifying \label{fig:SameLayerGen} Sketches of excluded lengths between a pair of particles on the same layer in (a) the same orientation and (b) different orientations. The excluded length is indicated by the red line measuring the smallest separation between the left-most points on the particles, marked by the vertical green lines. If they have opposite orientations the particle the excluded length is the width of the particles $w$. If they have the same orientation, they can get closer by an amount measured by $0<\alpha <1$.}
	\end{figure}

\begin{table}[]
    $W(\sigma_i,l_i;\sigma_{i+1},l_i) =$
\begin{tabular}{c||c|c}
\diagbox{$\sigma_{i+1}$}{$\sigma_{i}$} & +1 & -1\\    \hline \hline
+1 & $\alpha$ & $1$ \\    \hline
-1 & $1$ & $\alpha$ \\ 
\end{tabular}
  \captionof{table}{\justifying A table of the reduced excluded length $W$ for all possible combinations of $\sigma_i$ and $\sigma_{i+1}$ when the neighboring particles are on \textit{the same layer}. The parameter $\alpha$ must lie between zero and unity.}
    \label{tab:W0}
\end{table}
To provide an expression for $\mathcal{H}$, we need to return to the excluded lengths, $W(i,i+1)$. We first consider the case where the two particles are on the same layer, ${l_i = l_{i+1}}$. Here, the symmetries of the particle require $W$ to depend only on whether the particles are in the same orientation; i.e. the \textit{sign} of $\sigma_{i} \sigma_{i+1}$. When this is negative, $W = 1$ by definition, as shown in Fig.~\ref{fig:SameLayerGen}b. When it is positive, for the particles we consider, it must be \textit{smaller}; $W = \alpha$ with $0 \leq \alpha \leq 1$ (see Fig.~\ref{fig:SameLayerGen}a). The values of $W$ in this same layer case are compiled in Table \ref{tab:W0}. 

\begin{figure}\includegraphics[width=8cm]{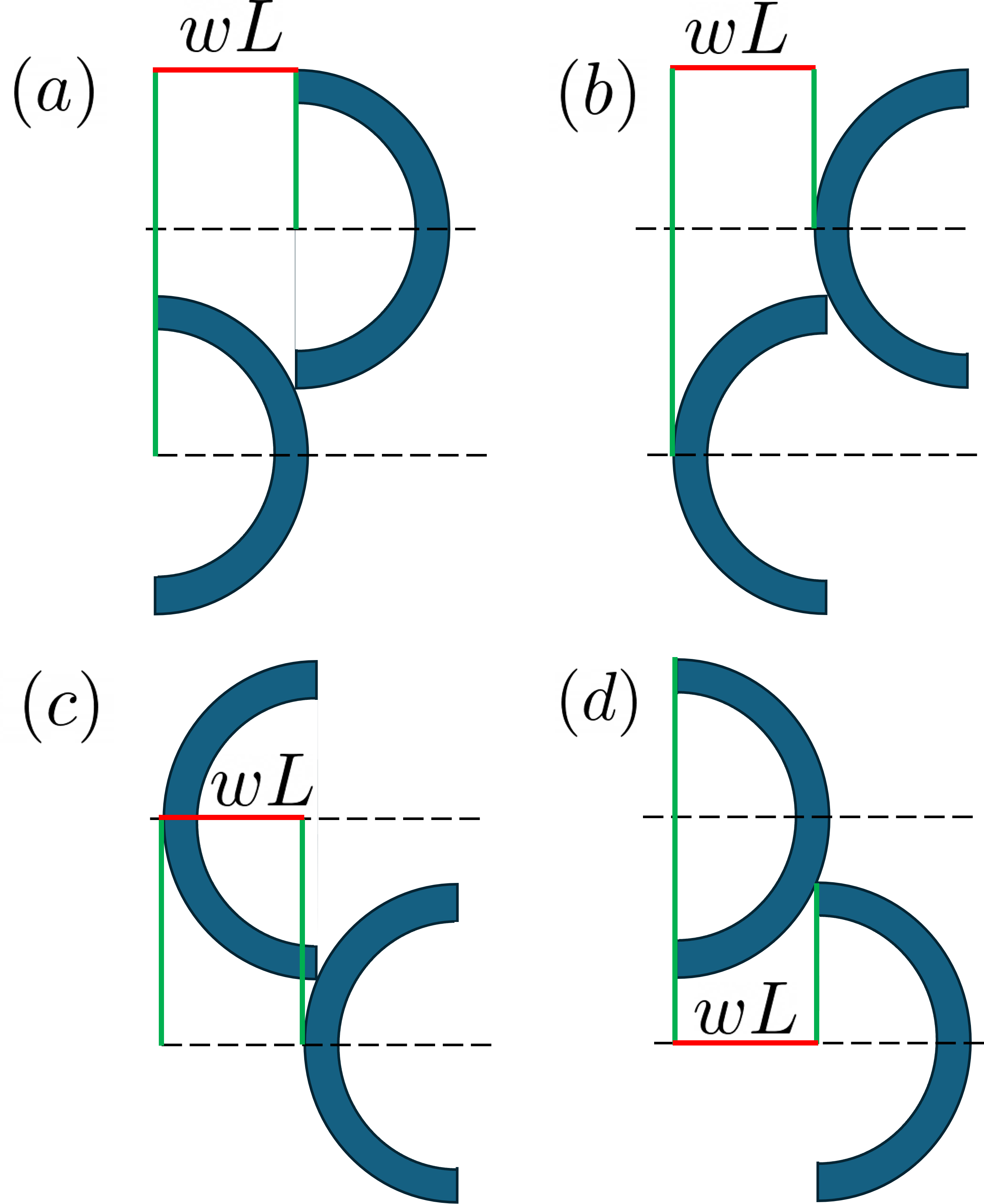}
		\captionof{figure}{\justifying \label{fig:DiffLayerGen} The excluded lengths between particles on different layers with the same orientations are all related by symmetry; rotating panel (a) half a turn clockwise will result in panel (b), reflecting it through a vertical line on the page results in panel (c) which is then rotated half a turn clockwise to obtain panel (d). The red line indicates the excluded length and is equal in each case to $w L$, which defines $L$.}
	\end{figure}

Next consider the case where the particle are on \textit{different} layers but with \textit{the same} orientations. Figure~\ref{fig:DiffLayerGen} shows that all four such configurations are related by symmetry via a series of rotations and reflections. Therefore, the dimensionless excluded length, $W$, must have \textit{the same value} in each of these cases. We shall call this $L$ and note that $0 \leq L \leq 1$.  
    
\begin{figure}\includegraphics[width=8cm]{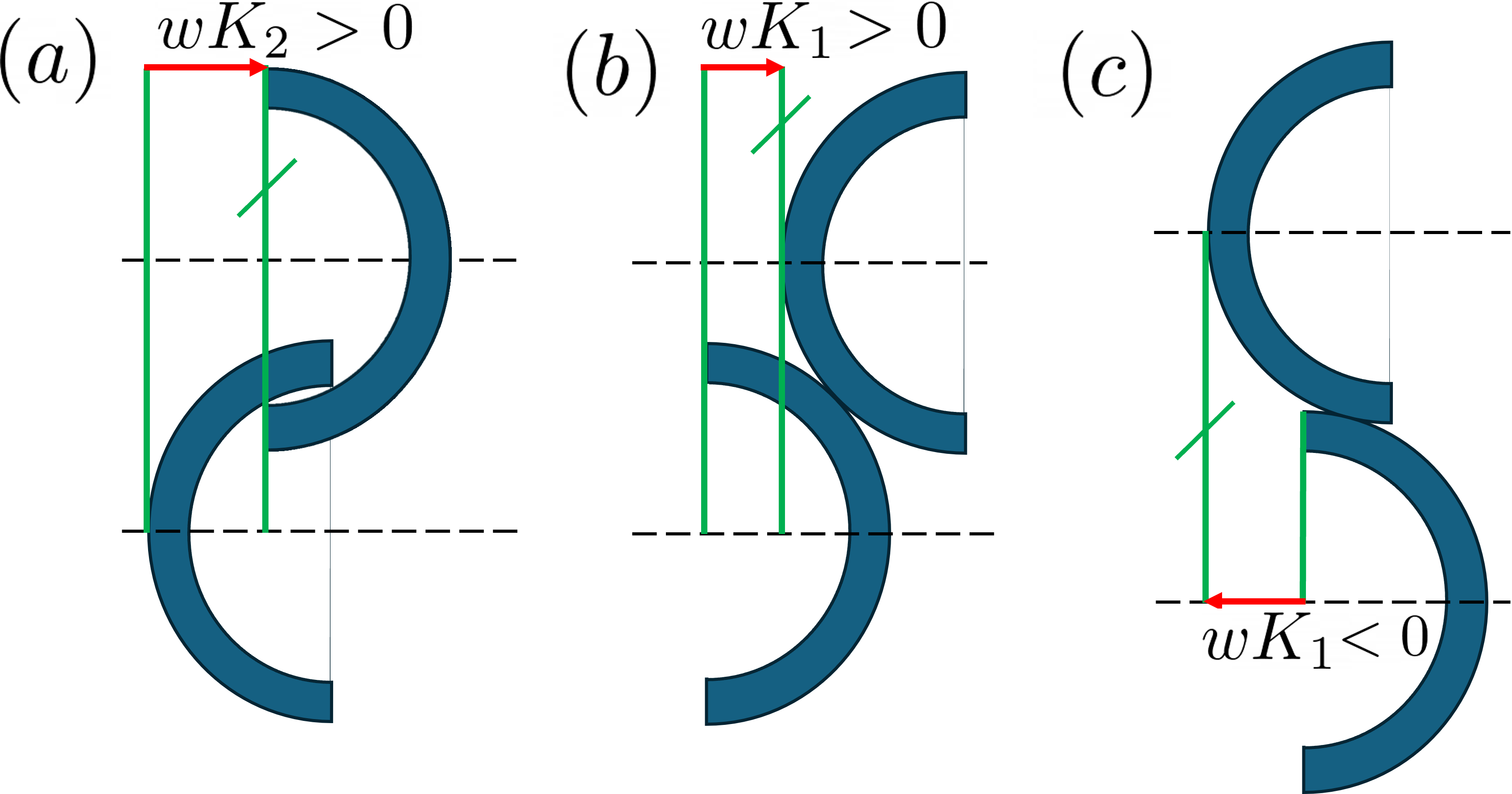}
		\captionof{figure}{\justifying \label{fig:K1K2Gen} Sketches showing the excluded lengths for particles on different layers in different orientations. Without loss of generality, we take $l_i = -1= -l_{i+1}$. In panel (a) we have $\sigma_{i}=-1=-\sigma_{i+1}$ and the lower particle is always on the left of the upper particle; the excluded length is always positive $w K_2 >0$. Panel (b) has $\sigma_{i}=+1=-\sigma_{i+1}$ and a small enough layer spacing that the lower particle stays on the left, $w K_1 >0$. If the layer spacing is increased, as in panel (c), then the lower particle can move to the right in this case making the excluded length negative, $w K_1 <0$. In each panel, the left-most points of the particles are shown by green vertical lines and that of the upper particle is marked with a tick. The excluded length is the distance between these lines and is shown by the red arrow; if the arrow points right it is positive, if it points left it is negative.}
	\end{figure}

Similar symmetries show us that $W$ in the two cases $l_i \neq l_{i+1}$ with $\sigma_{i}= + 1$ and $\sigma_{i+1} = -1$ are both equal to a value $K_1$, but generally \textit{different} from the pair $l_i \neq l_{i+1}$ with $\sigma_{i}= -1$ and $\sigma_{i+1} = +1$ whose $W$ value is $K_2$. It is critical to note that, while the latter must be \textit{positive}, $0 \leq K_2 \leq 1$, the former \textit{need not be}, $-1 \leq K_1 \leq 1$; as shown in Fig.~\ref{fig:K1K2Gen}. In Table~\ref{tab:Wh} we compile the values of $W$ when $l_{i} \neq l_{i+1}$.
\begin{table}[]
    $W(\sigma_i,l_{i},\sigma_{i+1},l_{i+1}) =$
\begin{tabular}{c||c|c}
\diagbox{$\sigma_{i+1}$}{$\sigma_{i}$} & +1 & -1\\    \hline \hline
+1 & $L$ & $K_2$ \\    \hline
-1 & $K_1$ & $L$\\ 
\end{tabular}
  \captionof{table}{\justifying A table of $W$ for all pairing of $\sigma_i$ and $\sigma_{i+1}$ when the neighboring particles are on \textit{different layers}. The parameters must lie in the following ranges; $0 \leq L \leq 1$, $-1 \leq K_1 \leq 1$, $0 \leq K_2 \leq 1$.}
    \label{tab:Wh}
\end{table}

Converting these tables for $W$ into functions of $\sigma$ and $l$ allows us to read off the effective Hamiltonian, up to unimportant constants,
\begin{equation}
\label{eq:HGen}
    \mathcal{H}(\textbf{S},\textbf{L}) = \sum_i \mathcal{J}(\sigma_{i},\sigma_{i+1}) l_i l_{i+1} - J \sigma_i \sigma_{i+1}.
\end{equation}
Here we have defined, ${J = 1 - \alpha - L + \bar{K}}$, and $\mathcal{J}(\sigma, \sigma')= (\alpha-1 - L + \bar{K}) \sigma \sigma' - \widetilde{K}(\sigma - \sigma') + (\alpha + 1 - L - \bar{K})$. In these expressions we have used ${\bar{K} = (K_1 + K_2 )/2}$ and ${\widetilde{K}=(K_1 - K_2)/2}$. 
    
Note the remarkably simple form of the Hamiltonian (\ref{eq:HGen}). From the point of view of the layer variables, $l_i$, it is a 1D Ising model with zero external field but with site dependent exchange couplings, given by $\mathcal{J}$, whose values depend on the orientation variables $\sigma_i$. This form is not unfamiliar, resembling famous generalisations of the conventional Ising model, such as the Edwards-Anderson spin glass (albeit with a specific choice of random exchange coupling) \cite{Edwards1975TheoryGlasses,Edwards1975} or the four-component Ashkin-Teller model \cite{Ashkin1943StatisticsComponents,Wu1976Ashkin-TellerProblem}. In fact, if it were not for the ${\sigma - \sigma'}$ term appearing in the exchange coupling $\mathcal{J}$ above, then our effective Hamiltonian would \textit{precisely} be the most general 1D Ashkin-Teller model. This connection is unsurprising, since the Ashkin-Teller model can be understood as a pair of Ising models coupled across two layers. We shall see, however, that the small change to $\mathcal{J}$ is important, and does affect the calculation of correlation functions. Nevertheless, it does not alter the simple Ising-like structure in~\eqref{eq:HGen}. This allows the sum over the layer variables to be taken \textit{exactly}, making it particularly easy to compute averages of quantities which depend only on the particles' orientations; a positive boon given our interest in their orientational order.
\section{Orientation Correlations}
\label{sec:OrCorr}
It is natural to calculate the equivalent of the total magnetisation for the orientation ``spins'', ${\mathcal{M}_{\sigma} = \left\langle \sum_{i} \sigma_{i} \right\rangle}$, as well as the correlation between the orientations of a pair of particles separated by $n$ others, $\mathcal{C}_{\sigma \sigma}(n) = \langle \sigma_{i} \sigma_{i+n} \rangle$. The angle brackets here indicate averaging over the effective Boltzmann distribution (\ref{eq:PBolt}). The first step is to sum over all the layer variables which is done using well known results for the 1D Ising model \cite{Chaikin1995PrinciplesPhysics}, though details may be found in Appendix \ref{app:OrrCorr}. The outcome is a product of terms, one for each particle. From the form of $\mathcal{J}$, we notice that the term for particle $i$ depends only on $\sigma_{i}$ and $\sigma_{i+1}$. This is important. It allows the result to be expressed as the exponential of \textit{another effective Ising Hamiltonian}, $\widetilde{\mathcal{H}}(\textbf{S})$, which includes \textit{only nearest neighbour interactions}.
    
Next we observe that the $\widetilde{\mathcal{H}}$ can have \textit{no effective external field}, i.e. no terms linear in $\sigma_{i}$. This can also be seen from $\mathcal{J}$, whose only linear term is proportional to the difference $\sigma_{i} - \sigma_{i+1}$. The Hamiltonian, $\widetilde{\mathcal{H}}$, may therefore depend only on \textit{the sum} of these differences. Given we are free to assume periodic boundary conditions in the thermodynamic limit, this must vanish. 

Therefore, averages of observables that depend only on the orientations, $\sigma$, may be taken against the Boltzmann weight, $e^{-\beta \widetilde{\mathcal{H}}(\textbf{S})}$, where the new effective Hamiltonian has the standard Ising form
\begin{equation}
\label{eq:Heff}
    \widetilde{\mathcal{H}}(\textbf{S}) =  \sum_{i} \widetilde{J} \ \sigma_i \sigma_{i+1},
\end{equation}
with effective exchange coupling $\widetilde{J}$ whose detailed form is provided in appendix \ref{app:OrrCorr}. This immediately leads us to the conclusion that \textit{the particles never acquire a preferred orientation}, $\mathcal{M}_{\sigma} = 0$. This should not be surprising; the system is still one-dimensional so, despite its other complexities, no spontaneous order can be expected. We may also read off the orientational correlations,
    \begin{equation}
    \label{eq:Csigsig}
        \mathcal{C}_{\sigma \sigma}(n) = \left(\tanh \beta \widetilde{J}\right)^n \equiv e^{- \lambda n}.
    \end{equation}
Here, we have defined the ``inverse orientational correlation length'',
\begin{equation}
\label{eq:lambdafull}
\begin{split}
    \lambda = &- \log \bigg[\tanh \bigg( \beta J \\
    &+ \frac{1}{4}\log \frac{\cosh^2[2\beta (\alpha - L)]}{\cosh[2\beta(K_1 -1)]\cosh[2\beta(K_2 -1)]} \bigg)\bigg]
    \end{split}
\end{equation}
The lack of spontaneous orientational ordering may lead us to expect this inverse length to be equally mundane; becoming zero only when the effective temperature is also. Fortunately, this expectation is not met and, depending on the particle geometry, there \textit{does exist} a critical $\beta$ where it \textit{changes discontinuously}; its imaginary part jumps suddenly from zero to $\pi$\footnote{The fact that $\lambda$ is given by the logarithm of a real quantity is why it can only pick up an imaginary part which is a multiple of $\pi$, and the correlation function can only oscillate with a period of one or two particles.}. This represents the orientational correlations swapping from ferromagnetic to \textit{anti}-ferromagnetic at a critical packing fraction, with neighboring particle preferring to assume \textit{opposite}, rather than equal orientations.
\subsection{Transition Condition}
\label{sec:TranCond}
The presence of this transition is marked by the effective exchange coupling $\widetilde{J}$ becoming \textit{negative}. From the definition of the inverse correlation length in (\ref{eq:Csigsig}), $\widetilde{J}$ is the argument of the hyperbolic tangent in (\ref{eq:lambdafull}) and it becomes negative when
\begin{equation}
\label{eq:antiferrocond}
    \beta J < \frac{1}{4}\log \frac{\cosh[2\beta(K_1 -1)]\cosh[2\beta(K_2 -1)]}{\cosh^2[2\beta (\alpha - L)]}.
\end{equation}
Here we demonstrate that, for some particle shapes, there can indeed be a $\beta$, and hence a packing fraction $\varphi$, above which this inequality is satisfied, but not below. This lets us formulate geometric criteria the particle must meet for this transition to be observed. We also show that there are a range of particle shapes which \textit{always} meet this condition and are anti-ferromagnetic for all packing densities.

First, let us suppose that $J > 0$. Now, we expand the right hand side of (\ref{eq:antiferrocond}) for small $\beta$, and find that its leading term is $\mathcal{O}(\beta^2)$. Since we have assumed the left hand side is positive, this implies that for a range of sufficiently small $\beta$ the ``anti-ferromagnetic condition'' (\ref{eq:antiferrocond}) is \textit{not met}. Now, we expand for large $\beta$. In this limit, the right hand side becomes ${\sim \beta (1 - \bar{K} - |\alpha - L|)}$. This allows (\ref{eq:antiferrocond}) to be satisfied at sufficiently large $\beta$ if the particle geometry has either
\begin{subequations}
\label{eq:genconds}
\begin{equation}
\label{eq:gencond1}
    \alpha > \bar{K}, \ \ \text{if} \ \ \alpha < L 
    \end{equation}
or    
    \begin{equation}
    \label{eq:gencond2}
    L > \bar{K}, \ \ \text{if} \ \ \alpha > L.
\end{equation}
These must be paired with the condition $J>0$, which can be written
\begin{equation}
\label{eq:gencond3}
    \bar{K}+1 >  \alpha + L. 
\end{equation}
\end{subequations}
If either of (\ref{eq:gencond1}) or (\ref{eq:gencond2}) is met simultaneously with (\ref{eq:gencond3}), then we have shown that there exists a $\beta$ above which the orientation correlations swap from ferromagnetic to antiferromagnetic. However, if neither are met, then the shape will always be ferromagnetic. 

If instead, we took $J < 0$, then we would find that the anti-ferromagnetic condition (\ref{eq:antiferrocond}) is \textit{always} satisfied. This is straightforwardly seen at small $\beta$, while at large $\beta$ the equivalents to (\ref{eq:gencond1}) are (\ref{eq:gencond2}) are: $\alpha < 1$ and $L < 1$, both of which are true by definition. Therefore, all particles which \textit{do not} satisfy (\ref{eq:gencond3}) are always anti-ferromagnetic. 

This discussion raises the natural question: ``What do the particles exhibiting these behaviors look like?''

\section{Going Bananas: particles that flip}
\label{sec:Bans}
Until now, our discussion has been quite general, and our results may be applied to any particle shapes which satisfy the conditions in section \ref{sec:System}. Here we shall construct a one-parameter family of particle shapes which, as a function of that parameter, exhibit the full range of possible behaviours for their orientation correlations.

\begin{figure}\includegraphics[width=8.4cm]{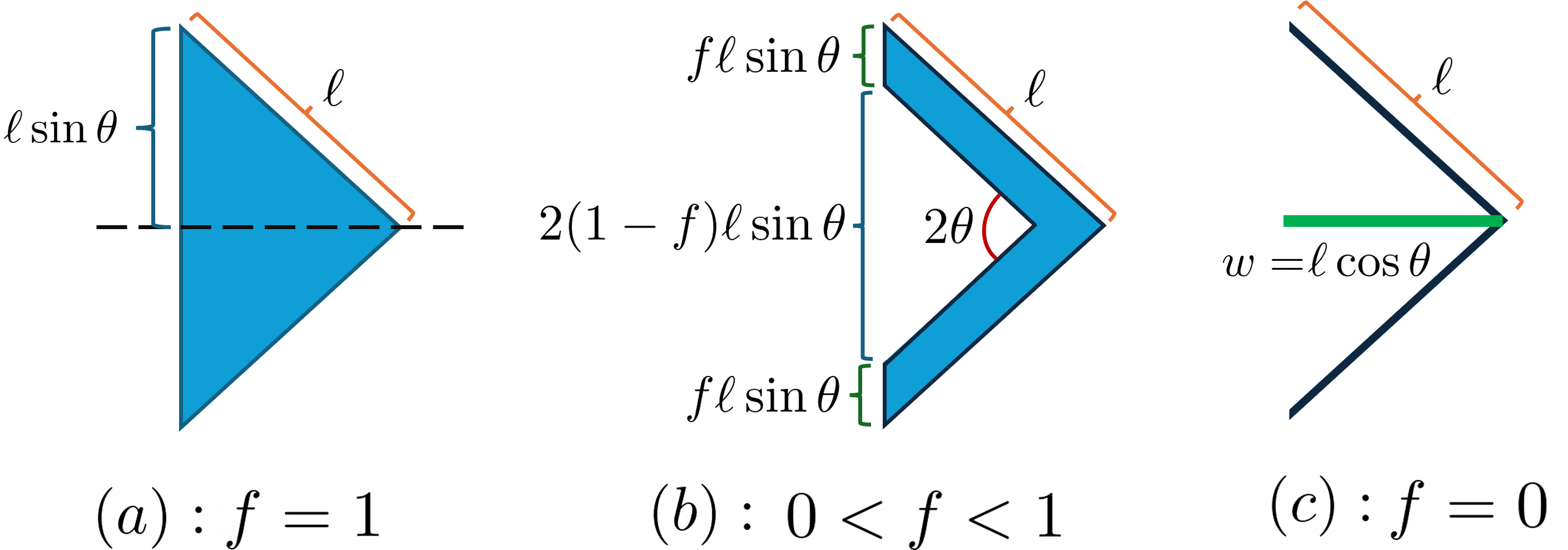}
		\captionof{figure}{\justifying \label{fig:Lshaped} Sketches of L-shaped banana particles we consider. This family of shapes is made by taking ever larger bites, measured by $f$, out of the isoceles triangle shown in panel (a) with equal sides of length $\ell$ and opening angle $2 \theta$. Panel (b) shows the geometry for an intermediate bite size $0<f<1$ and panel (c) shows the mathematically thin L-shape defined as the limit $f=0$. In each case the width of the particle is $w = \ell \cos \theta$.}
	\end{figure}

We start with a slice of cake; an Isosceles triangle whose equal sides have length $\ell$ and whose apex angle is $2 \theta$, shown in Fig.~\ref{fig:Lshaped}. The layer bisects this angle so that the unequal side is oriented vertically. Then we take a bite; remove a similar Isosceles triangle, identical in dimensions but smaller by a factor $1 - f$ than the original, from the unequal side to result in an L-shaped ``banana'', as shown in Fig.~\ref{fig:Lshaped}b. As the size of the bite is varied, we produce a family ranging from a full slice of banana cake, $f = 1$ in Fig.~\ref{fig:Lshaped}a, to a mathematically thin L-shape which we define as the limit when $f \to 0$ in Fig.~\ref{fig:Lshaped}c. 

\begin{figure}\includegraphics[width=8cm]{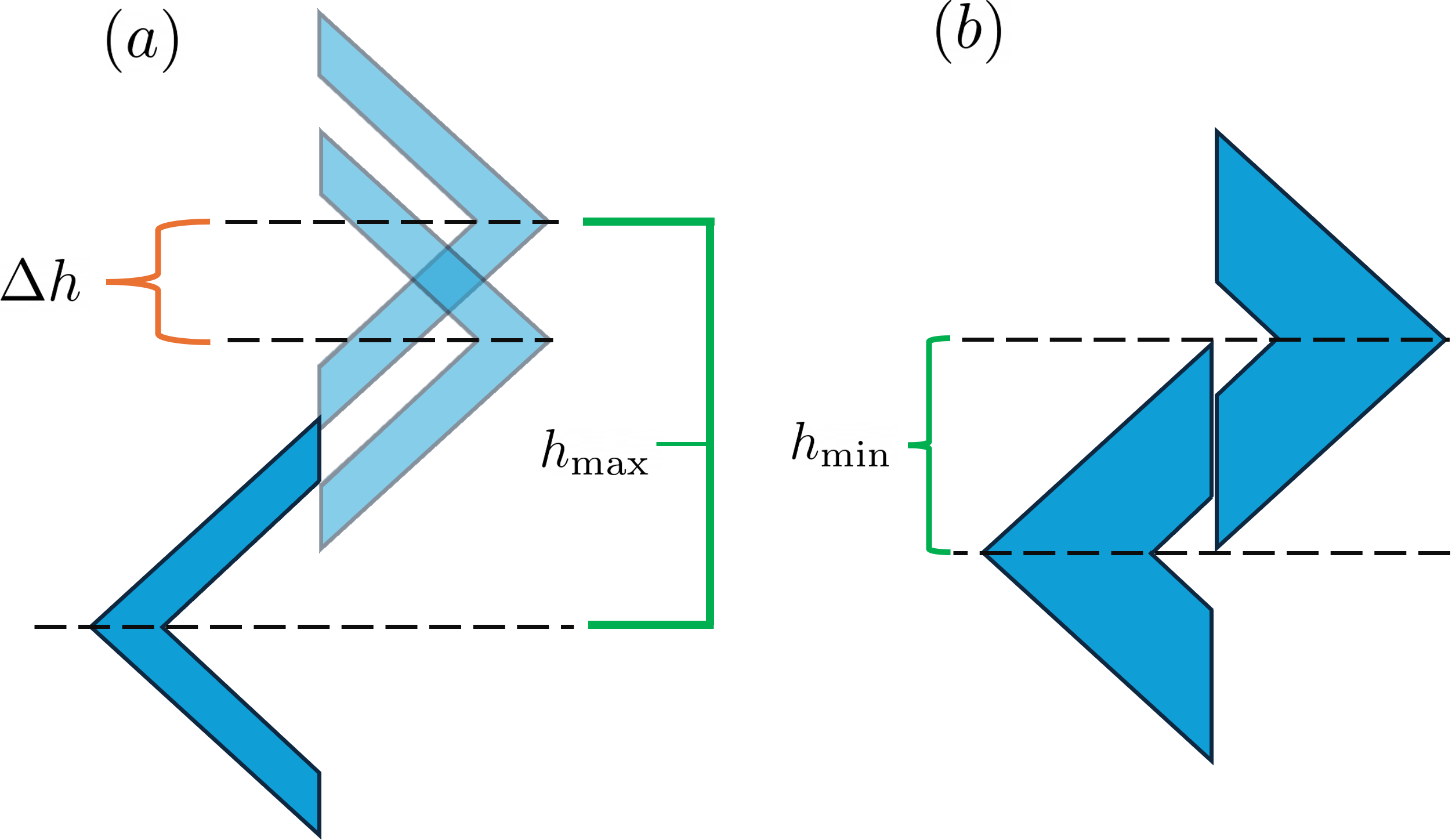}
		\captionof{figure}{\justifying \label{fig:BigBites} Sketches showing the effect of the bite size and layer spacing on $K_2$. Panel (a) shows that for any bite size $0<f<1$ there is a range of large layer spacings, from $h_{\text{max}}$ to $h_{\text{max}}-\Delta h$, for which the particles exclude as much length as full triangles. Panel (b) shows that if the bites are sufficiently small, $\Delta h$ is large enough that no difference from full triangles is seen.}
	\end{figure}

It is an exercise in planar geometry to determine the excluded widths as a function of the bite size, $f$, and the layer spacing $h$. Leaving the details to appendix \ref{app:Geo}, we begin by defining the convenient parameter
\begin{equation}
\label{eq:gdef}
    g = \frac{h}{\ell \sin \theta} - 1,
\end{equation}
which ranges between $0 < g < 1$ and measures the ratio of the layer spacing to the particle size. In terms of this and $f$, we find: 
\begin{subequations}
\label{eq:Constants}
    \begin{equation}
        \alpha = f,  \ \ \ L = 1 - g, \ \ \text{and} \ \ K_1 = - g.
    \end{equation}
The situation is more complicated for $K_2$, as shown in Fig.~\ref{fig:BigBites}. For a finite bite size, there exists a range of sufficiently large layer spacings where it is \textit{not} noticed and $K_2 = 1$ as it was for the complete triangle, as shown in Fig.~\ref{fig:BigBites}a. Only reducing the layer spacing sufficiently sees $K_2$ begin to decrease. In fact, for small enough nibbles, no difference from triangles is ever seen, as in Fig.~\ref{fig:BigBites}b. The piecewise result is,
    \begin{equation}
    \label{eq:K2}
K_2 =\begin{cases}
			1, & \text{if $g > 1 - 2 f$}\\
            g + 2 f, & \text{if $g < 1 - 2 f$.}
		 \end{cases}
    \end{equation} 
\end{subequations}
Notice that the first condition is \textit{always} met if $f > 1/2$, in which case $K_2 = 1$, no matter the value of $g$.

Depending on the size of the bite, $f$, these particles can exhibit the full range of behaviours for their orientational correlations. Using the definitions (\ref{eq:Constants}) in the criteria (\ref{eq:genconds}) we can quickly ascertain that if the bite is quite small, $f > 1/2$, then only for large layer spacings with ${g > 2 f - 1}$ will the antiferromagnetic transition be seen. For smaller layer spacings, these particles will be \textit{antiferromagnetic for all packing fractions}. 

The large bite case, $0 < f < 1/2$, warrants slightly closer inspection. These particles all meet condition (\ref{eq:gencond3}) so, to see the antiferromagnetic transition, they must meet either (\ref{eq:gencond1}) or (\ref{eq:gencond2}). If the layer spacing is large enough, $g > 1 - f$, then (\ref{eq:gencond2}) is the relevant condition and, from the definition of $K_2$ (\ref{eq:K2}), it is easy to see that it is \textit{always} satisfied; the antiferromagnetic transition is seen. When $g < 1 - f$, we need to satisfy (\ref{eq:gencond1}). Again, it is straightforward to show that this is true when $g > 1 - 2f$. However, if $g < 1 - 2f$, the situation appears inconclusive; $\alpha = \bar{K}$ \textit{precisely}. This is because the first order term in the expansion of the right hand side of the transition condition (\ref{eq:antiferrocond}) for $\beta \gg 1$ \textit{exactly cancels} the left hand side. The next order in the expansion of the right hand side is \textit{positive} and so the transition occurs, but its quality is different. 

Consider $1/\Re (\lambda)$. This is understood as the orientational correlation length and measures the distance over which the orientational order decays, be it ferromagnetic or antiferromagnetic. Expanding for $\beta \gg 1$ when $g < 1 - 2f$ we find that $1/\Re (\lambda) \propto \beta^{-1}$. This becomes extremely small at high packing fractions and is in contrast to the growing case when $g > 1 - 2f$ where $1/\Re (\lambda) \propto e^{k \beta}$ with $k > 0$. This means that, while the antiferromagnetic transition is seen for large bites $0 < f < 1/2$, when the layer spacing is small, $g < 1 - 2f$, the transition is somewhat ``weakened'' and the degree of antiferromagnetic ordering decreases, becoming very small for high densities.  

Finally, note the end cases. The full triangle, with $f=1$, is always antiferromagnetic while the thin L-shape, with $f=0$, can be shown to be \textit{ferromagnetic}, no matter the density\footnote{This ultimately stems from the fact that $\alpha = 0$ and $1 - K_2 = L$, which causes the $K_2$ factor on the right hand side of (\ref{eq:antiferrocond}) to cancel. This was the factor which led to the negative term in the large $\beta$ expansion. In its absence, this term is positive.}. 
\begin{figure}\includegraphics[width=8cm]{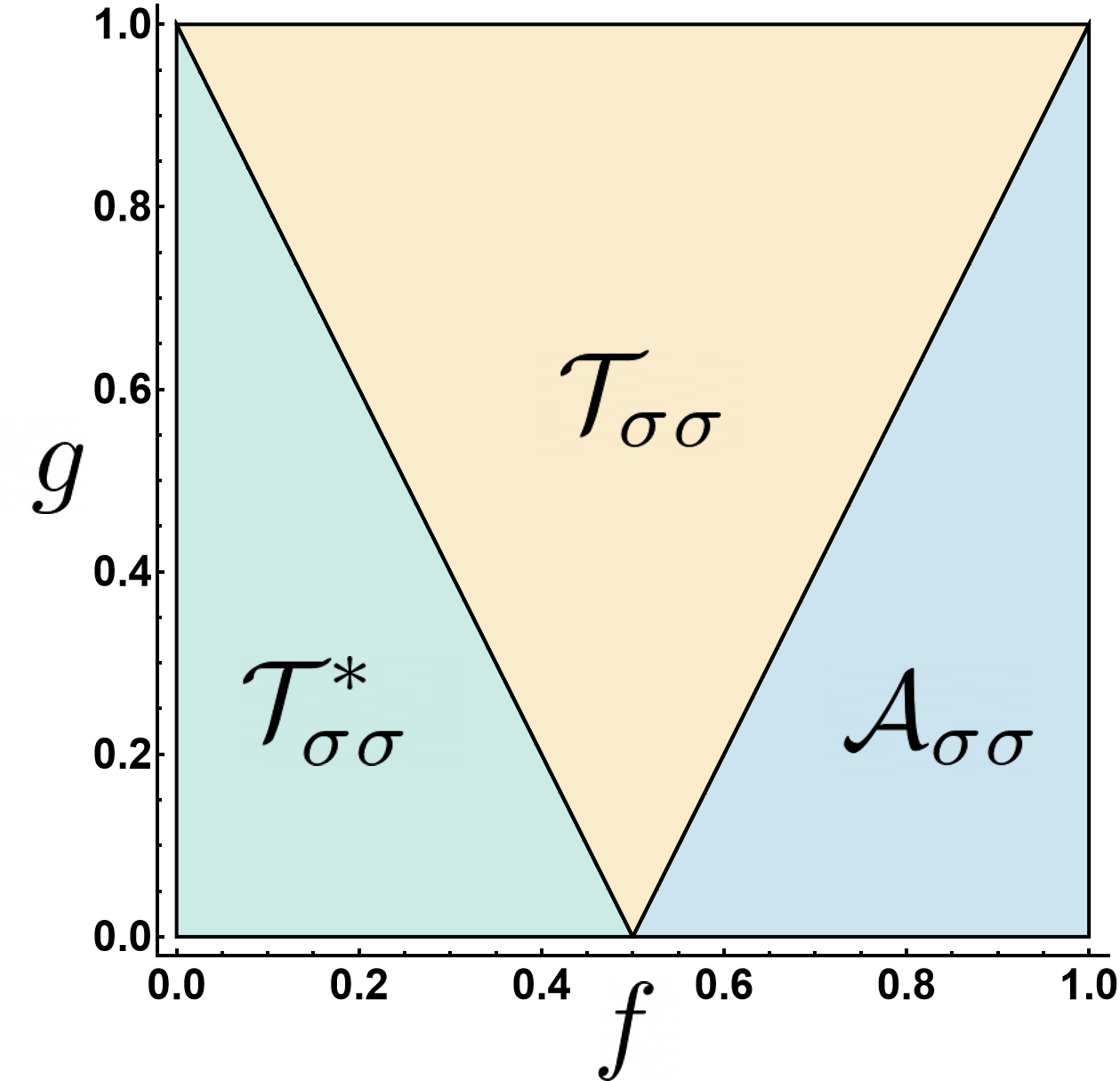}
		\captionof{figure}{\justifying \label{fig:OrrOrd} The behaviour of the orientational correlation function, $\mathcal{C}_{\sigma \sigma}$, for different layer spacings, $g$, and bite sizes $f$. In the blue region on the far right, the order is totally antiferromagnetic and is labelled $\mathcal{A}_{\sigma \sigma}$. In the central yellow region, labelled $\mathcal{T}_{\sigma \sigma}$, the correlations transition from ferromagnetic and low densities to antiferromagnetic at high densities, where the correlation length diverges. This transition is weakened in the the green region on the far left with the correlation length vanishing at high densities. We distinguish this from the central region with the symbol $\mathcal{T}^{*}_{\sigma \sigma}$.}
	\end{figure} 

    \begin{figure*}\includegraphics[width=\textwidth]{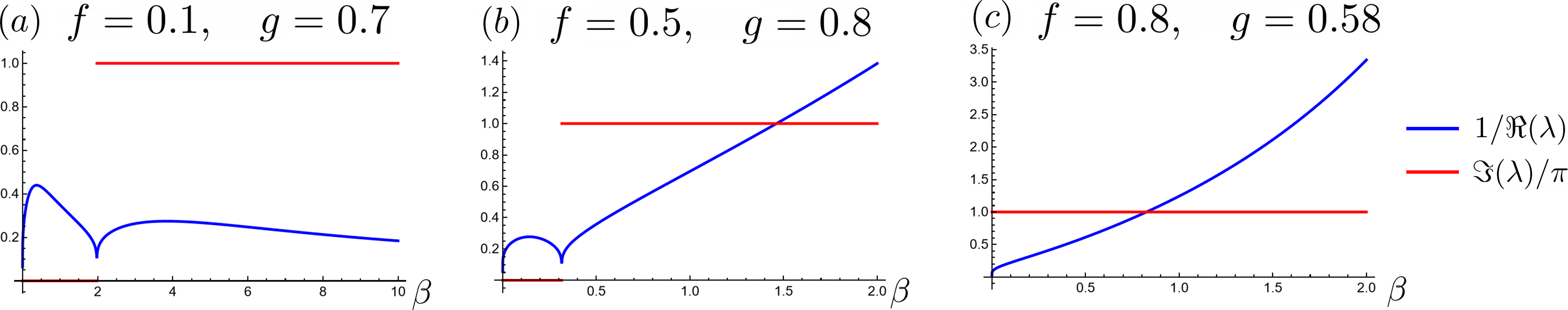}
		\captionof{figure}{\justifying \label{fig:CssCorrLengths} Plots of the correlation length, $1/\Re(\lambda)$ shown in blue, and the oscillation period $\Im(\lambda)/\pi$ of the orientational correlation function $\mathcal{C}_{\sigma \sigma}(n)$ as a function of packing density, $\beta$, for various different bite sizes, $f$, and layer spacings $g$. In panel (a) the system is in the $\mathcal{T}^{*}_{\sigma \sigma}$ region of Fig.~\ref{fig:OrrOrd} and the correlations are initially ferromagnetic. As the packing density is increased, $\Im(\lambda)$ discontinuously becomes non-zero, indicating a transition to antiferromagnetic correlations. Precisely at the transition point, the correlation length decreases sharply to zero. Here, as the packing approaches its maximal density ($\beta \to \infty$) the correlation length decreases. In panel (b) the system is in the central $\mathcal{T}_{\sigma \sigma}$ region, and once again the transition from ferromagnetic to antiferromagnetic ordering is seen. In this case, however, the correlation length diverges at the densest packings. Finally in panel (c) the system is in $\mathcal{A}_{\sigma \sigma}$ so the correlations are always antiferromagnetic, and this ordering gets stronger as packing density increases.} 
	\end{figure*} 

The behaviour of the the orientational correlations for different values of the parameters $f$ and $g$ is summarised in Fig.~\ref{fig:OrrOrd}. We have labelled the regions on this plot as follows: $\mathcal{A}_{\sigma \sigma}$ stands for totally antiferromagnetic where, no matter the density, the orientational order is always antiferromagnetic. The symbol $\mathcal{T}_{\sigma \sigma}$ means there is a transition from ferromagnetic to antiferromagnetic ordering as the packing fraction increases to its maximum value, where the correlation length of the antiferromagnetic ordering diverges. In the region labelled $\mathcal{T}^{*}_{\sigma \sigma}$, the weakened antiferromagnetic transition is seen where, at the highest packing fractions, the correlation length approaches zero. 

In Fig.~\ref{fig:CssCorrLengths}, we plot the correlation length $1/\Re(\lambda)$ (shown in blue) and the oscillation period $\Im(\lambda)/\pi$ (shown in red) in each of the three regions of the ordering diagram, Fig.~\ref{fig:OrrOrd} as a function of packing fraction through $\beta$. The transitions from ferromagnetic to antiferromagnetic orientational ordering are clearly seen when the oscillation period jumps from zero to unity as the packing fraction is increased. This is accompanied by the correlation length rapidly approaching zero at the transition point. After which it may either monotonically increase or drop back to zero at high densities.

\section{Layer-Layer \& Layer-Orientation Correlations}   
\label{sec:LLCorr}
To understand the layering order of the particles we must, of course, understand the variables, $l_i$. It is also natural to consider the combination $\sigma_i l_i$, whose sign probes the preference for those particles on a given layer to have the same orientation, analogous to the layer polarisation in the $\text{SmA-P}$ phases discussed previously. 

As was the case for the orientations, we cannot expect to see any spontaneous order in these variables. Indeed both ``magnetisations'' vanish, ${\langle \sum_{i} l_i \rangle = \langle \sum_{i} \sigma_i l_i \rangle = 0}$, by virtue of the system being one dimensional \footnote{This is also seen more concretely by noting that the system is symmetric when reflected about the line bisecting the two layers. As a result the effective Hamiltonian (\ref{eq:HGen}) is symmetric under changing the signs of all $l_i$ and averages containing an \textit{odd} number of $l_i$s vanish}. On the other hand, as we have just seen, the correlations need not be so simple. 

Let us consider the layer-layer correlations, ${\mathcal{C}_{ll}(n) = \langle l_i l_{i+n} \rangle}$, and the layer-orientation cross correlations, ${\mathcal{C}_{\sigma l}(n) = \langle \sigma_i l_i \sigma_{i + n} l_{i+n} \rangle}$. These will be more difficult to calculate than $\mathcal{C}_{\sigma \sigma}$, because the appearance of the $l_i$ inside the average. The full calculation is reported in appendix \ref{app:Corr}, but the scheme is as follows. First, we define the function ${\mathcal{G}(\bm{\mu},\bm{\xi}) = -\beta^{-1} \sum_{i} (\mu_i l_i l_{i+1} + \xi_i \sigma_i l_i \sigma_{i+1} l_{i+1})}$, which depends on the introduced variables, ${\bm{\mu} = (\mu_1, \cdots, \mu_{N})}$ and ${\bm{\xi} = (\xi_1, \cdots, \xi_{N})}$. This is added to the effective Hamiltonian (\ref{eq:HGen}), resulting in the associated partition function $\mathcal{Z}(\bm{\mu},\bm{\xi})$. The correlation functions are then found in the usual way by taking the appropriate derivatives of the partition function, for example 
\begin{equation}
    \label{eq:CorrDiffs}
    \mathcal{C}_{ll}(n) = \frac{1}{\mathcal{Z}}\left(\prod_{j=i}^{i+n-1}\frac{\partial}{\partial \mu_{j}}\right) \mathcal{Z}(\bm{\mu},\bm{\xi})\bigg\lvert_{\bm{\mu} = \bm{\xi}= \bm{0}}.
\end{equation}

This introduction of $\bm{\mu}$ and $\bm{\xi}$ just changes the constants in $\mathcal{J}$ but does not change the structure of $\mathcal{H}$. This means that the sum on $\textbf{L}$ may be taken just as before. The difference is that the new effective Hamiltonian, $\widetilde{\mathcal{{H}}}(\bm{\mu},\bm{\xi})$, that results will depend on the introduced variables in a protracted way. This prevents us from computing the correlation functions by direct analogy to known results for the 1D Ising model because, when the derivatives with respect to these variables are taken as in (\ref{eq:CorrDiffs}), we end up needing to compute averages of elaborate and \textit{non-local} combinations of $\sigma_i$s. Whereas $\mathcal{C}_{\sigma \sigma}$ only depended on the pair of orientations $\sigma_{i}$ and $\sigma_{n}$, both $\mathcal{C}_{ll}$ and $\mathcal{C}_{\sigma l}$ depend on these \textit{and all other orientations in between}. This is a reflection of the complicated interplay between the particles' layering and orientational order and results in more detailed correlation functions. 

These correlation functions compare, in some sense, the length excluded when the particles are on the same or different layers. As such, from Tables \ref{tab:W0} \& \ref{tab:Wh}, they naturally depend on the differences $(\alpha - L)$, $(1 - K_1)$ and $(1- K_2)$ \footnote{This is also true of $\mathcal{C}_{\sigma \sigma}$ as is seen in (\ref{eq:Csigsig}) \& (\ref{eq:lambdafull}).}. In particular, they are sensitive to the cases when any one of these differences is zero. For the banana particles we have considered, this is only relevant for $K_2$ which, as seen in (\ref{eq:K2}), is exactly unity for a range of layer spacings and particle geometries. 

The cases $K_2 = 1$ and $K_2 \neq 1$ each require a different calculation strategy (found in appendix \ref{app:Corr}) and result in different forms of the correlations. When $K_2 = 1$ we find
\begin{subequations}
\label{eq:CK2eq1}
\begin{equation}
\label{eq:CK2eq1ll}
    \mathcal{C}_{ll}(n) = e^{-n \kappa} \left(1- n A\right),
\end{equation}
and
\begin{equation}
\label{eq:CK2eq1sl}
     \mathcal{C}_{\sigma l}(n) = e^{-n \kappa} \left(1+ n A\right).
\end{equation}
\end{subequations}
Here we have defined the ``inverse decay length''
\begin{equation}
\label{eq:kappa}
    \kappa = \log \left(\frac{\cosh \beta \widetilde{J}}{\tanh[2\beta(L - \alpha)]}\right) - \beta \widetilde{J},
\end{equation}
in terms of the effective coupling constant $\widetilde{J}$ appearing in the orientational correlation function (\ref{eq:Csigsig}) and (\ref{eq:lambdafull}). The constant $A$ is defined by
\begin{equation}
\label{eq:Adef}
    A = \frac{\tanh[2\beta (1- K_1)]}{2\tanh[2\beta(L - \alpha)]} e^{- 2\beta\widetilde{J}}.
\end{equation}

If, on the other hand $K_2 \neq 1$, the results are
\begin{subequations}
\label{eq:CK2neq1}
\begin{equation}
    \mathcal{C}_{ll}(n) = \cosh^2(y) e^{-n \eta} - \sinh^2(y) e^{-n \nu},
\end{equation}
and
\begin{equation}
    \mathcal{C}_{\sigma l}(n) = \cosh^2(y) e^{-n \nu} - \sinh^2(y) e^{-n \eta},
\end{equation}
\end{subequations}
where we have defined 
\begin{equation}
\label{eq:ydef}
    y = \frac{1}{4} \log \frac{\tanh[2\beta(1-K_1)]}{\tanh[2\beta(1-K_2)]},
\end{equation}
which makes plain that these formulae are only useful when $K_2 \neq 1$. The inverse decay lengths, $\eta$ and $\nu$, appearing in the exponentials above have hopelessly complicated expressions which are reported in appendix \ref{app:LLCorr}. For the purposes of our discussion, their relevant properties will simply be stated and the interested reader is directed to appendix \ref{app:etamu} for their proofs. 

Before moving to discuss the behaviour of these correlation functions for the banana particles, pause to note their general form. Unlike, the orientational correlations (\ref{eq:Csigsig}), neither $\mathcal{C}_{ll}$ nor $\mathcal{C}_{\sigma l}$ are expressed as a single exponential (see eqs. (\ref{eq:CK2eq1}) \& (\ref{eq:CK2neq1})). This is another demonstration of the complex coupling between orientational and layer ordering. 
\subsection{Application to Bananas}
\label{sec:LLCorrBan}
Let us apply these expressions to the banana particles introduced in Sec.~\ref{sec:Bans}. There, in Fig.~\ref{fig:OrrOrd}, we presented a diagram which showing the orientational ordering as a function of the layer spacing $g$ and bite size $f$. The goal of this section is to complete that diagram with the behaviours of $\mathcal{C}_{ll}$ and $\mathcal{C}_{\sigma l}$. We will break our discussion up according to the three regions in Fig.~\ref{fig:OrrOrd}

\subsubsection{\texorpdfstring{$\mathcal{T}^{*}_{\sigma\sigma}$: $g< 1- 2f$}{T*}}
\label{sec:Tss}
We begin in the bottom left region, where $g<1 - 2f$. The relevant correlations here are (\ref{eq:CK2neq1}) and, just as for the orientation correlations, our goal is to understand their behaviour at high and low packing fractions. 

We first turn our attention to dense packings and observe, from (\ref{eq:ydef}), that  $y\to 0$ as $\beta \to \infty$. This means that the decay of $\mathcal{C}_{ll}$ at high densities is controlled by $\eta$, and its brother $\mathcal{C}_{\sigma l}$ subsides according to $\nu$. Calling upon our analysis in appendix \ref{app:etamu}, we find that, for large $\beta$, one is \textit{very large}, $\eta \propto \beta$, while the other is \textit{minute}, $\nu \propto e^{-b \beta}$, with $b>0$. This shows us that in this region at the highest packing fractions there is \textit{very little layering order} ($\mathcal{C}_{ll} \to 0$) but there is \textit{significant correlation between} the layers sat on by the particles and their orientations ($|\mathcal{C}_{\sigma l}| \to 1$). We may go further. At large $\beta$, both $\eta$ and $\nu$ are \textit{real}. Thus, these correlations are \textit{ferromagnetic} for dense packings. 

\begin{figure}\includegraphics[width=8.5cm]{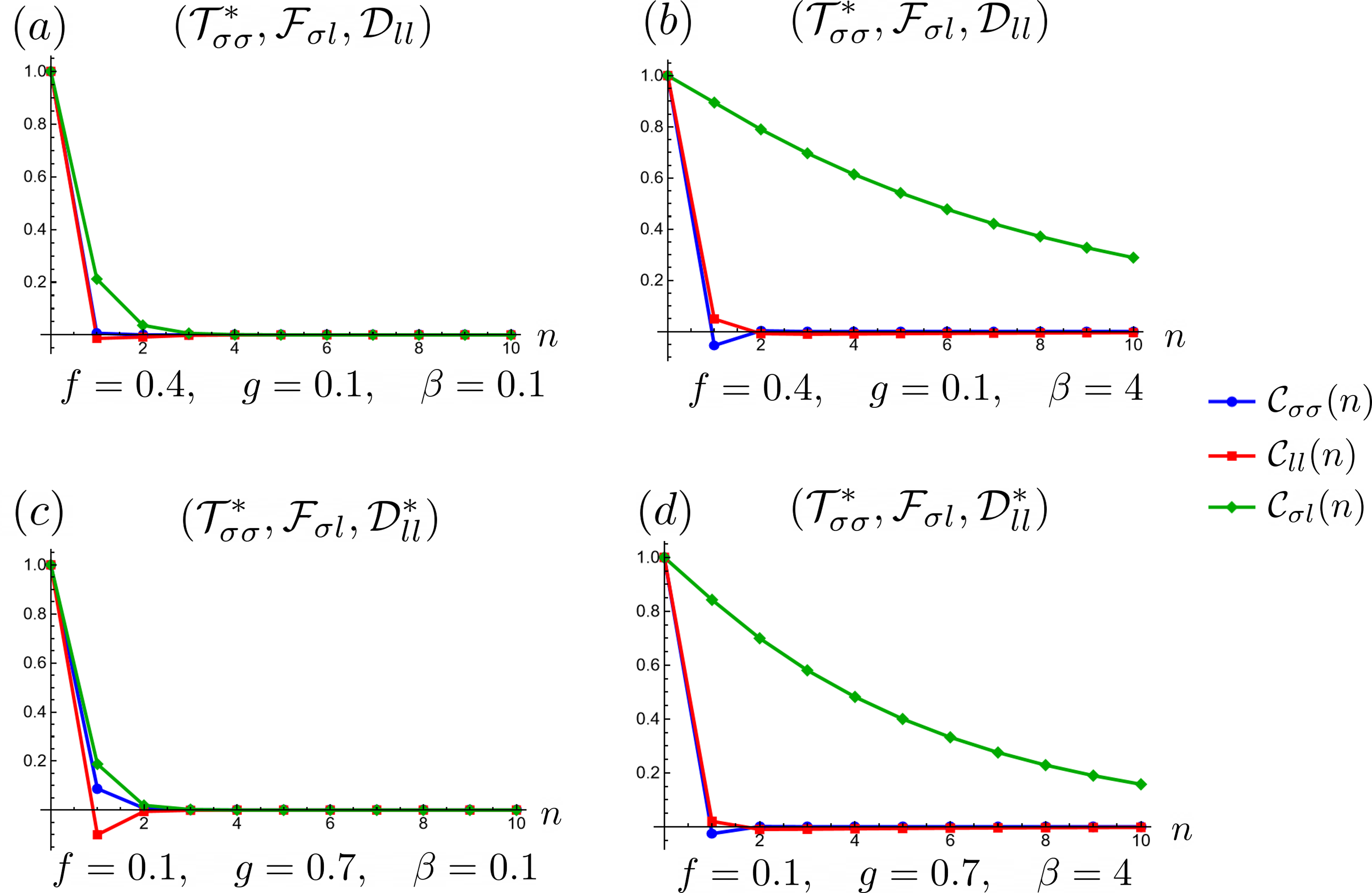}
		\captionof{figure}{\justifying \label{fig:CorrsLeft} Typical behaviour of the correlation functions in the $(\mathcal{T}^{*}_{\sigma \sigma},\mathcal{F}_{\sigma l},\mathcal{D}_{l l})$ and $(\mathcal{T}^{*}_{\sigma \sigma},\mathcal{F}_{\sigma l},\mathcal{D}^*_{l l})$ regions of Fig.~\ref{fig:AllOrd}. Panels (a) and (c) are for low packing densities while (b) and (d) are for high densities. In (a) and (b), the layer-orientation correlations, shown in green, are ferromagnetic, with the degree of ordering increasing as packing density increases. The orientational correlations, shown in blue, transition from ferromagnetic to antiferromagnetic, but the correlation length of the order remains small. The layer-layer correlations, shown in red, can become negative, although they do not have a consistent sign pattern. This is difficult to see, given the very short correlation length; these correlations are essentially disordered. The behaviours of $\mathcal{C}_{\sigma \sigma}$ and $\mathcal{C}_{\sigma l}$ in (c) and (d) are the same as in (a) and (b), but $\mathcal{C}_{ll}$ is subtly different. It now has an oscillatory component, making its sign change more pronounced at low densities (compare its value at $n=1$ in panels (a) and (c)). } 
	\end{figure}

For lower densities the situation is less clear-cut. Here $y$ approaches a constant, meaning each decay in $\mathcal{C}_{ll}$ and $\mathcal{C}_{\sigma l}$ carries equal weight. The behaviour of $\nu$ is easy to understand; it is real, positive and large $\nu \sim |\log \beta|$. The waters are muddied by $\eta$. It is also large $|\eta| \sim |\log \beta|$, but it may be shown, by a similar argument to that in section \ref{sec:TranCond} (see appendix \ref{app:etamu}), if $(\alpha - L)^2 < (1- K_1)(1-K_2)$, it has an imaginary part of $i \pi$. Thus, we might expect oscillations in the correlation functions. However, since both correlation lengths are very small ($\sim |\log \beta|^{-1}$), and the oscillations from $\eta$ need to overcome the pure decay from $\nu$, $\mathcal{C}_{ll}$ and $\mathcal{C}_{\sigma l}$ both decay very quickly for low densities reaching essentially zero for $n \gtrsim 2$; only $\mathcal{C}_{ll}$ manages to change sign noticeably. Example behaviours of the correlation functions are shown in Fig.~\ref{fig:CorrsLeft}.

To summarise, at low densities the particles choose random orientations and are distributed almost randomly between layers with only nearest neighbours preferring to be on opposite layers. At high densities, the lack of layering order persists but those particles who share a layer are likely to also share the same orientation; the layers become polarised. Therefore, we label this region with $\mathcal{F}_{\sigma l}$, indicating that the correlation between the orientations and layers is ferromagnetic with the degree of ordering increasing for increasing packing fraction. For the layering order we give the label $\mathcal{D}_{ll}$, to indicate it is essentially disordered, when $\Im( \eta) = 0$ and $\mathcal{D}^*_{ll}$, when $\Im(\eta) = i \pi$. The star here indicates that $\mathcal{C}_{ll}$ may become noticeably negative for small $n$.
\subsubsection{\texorpdfstring{$\mathcal{T}_{\sigma\sigma}$: $1-2f<g> 2f-1$}{T}}
We now move to the central region of Fig.~\ref{fig:OrrOrd}. Here ${K_2 =1}$ so we must use (\ref{eq:CK2eq1}) for the correlations. Evidently, the decay of both correlations is controlled by $\kappa$. From its expression (\ref{eq:kappa}), the only way this can pick up an imaginary part is if the argument of its hyperbolic tangent becomes negative. Hence, if $g> 1 - f$ both $\mathcal{C}_{ll}$ and $\mathcal{C}_{\sigma l}$ show \textit{antiferromagnetic} behaviour, but are ferromagnetic otherwise. It is straightforward to show that at high packing fractions $\kappa \sim \beta \gg 1$, and so the degree of this ordering is weak. 

The apparently simple ferromagnetic order does warrant a closer look. Inspecting (\ref{eq:CK2eq1}) we see that, for large $n$, $\mathcal{C}_{ll}$ can be \textit{negative} as long as $A$ is positive, which it is precisely when $g<1-f$. Hence, the layer-layer correlations, $\mathcal{C}_{ll}$, are in fact \textit{anti-correlations} here; no particles ever want to share their layer. Note that $\mathcal{C}_{\sigma l}$ is still positive here. It can, in principle, change sign when $A$ does; however, this is when both $\mathcal{C}_{ll}$ and $\mathcal{C}_{\sigma l}$ are anti-ferromagnetic with no fixed sign.

\begin{figure}\includegraphics[width=8.5cm]{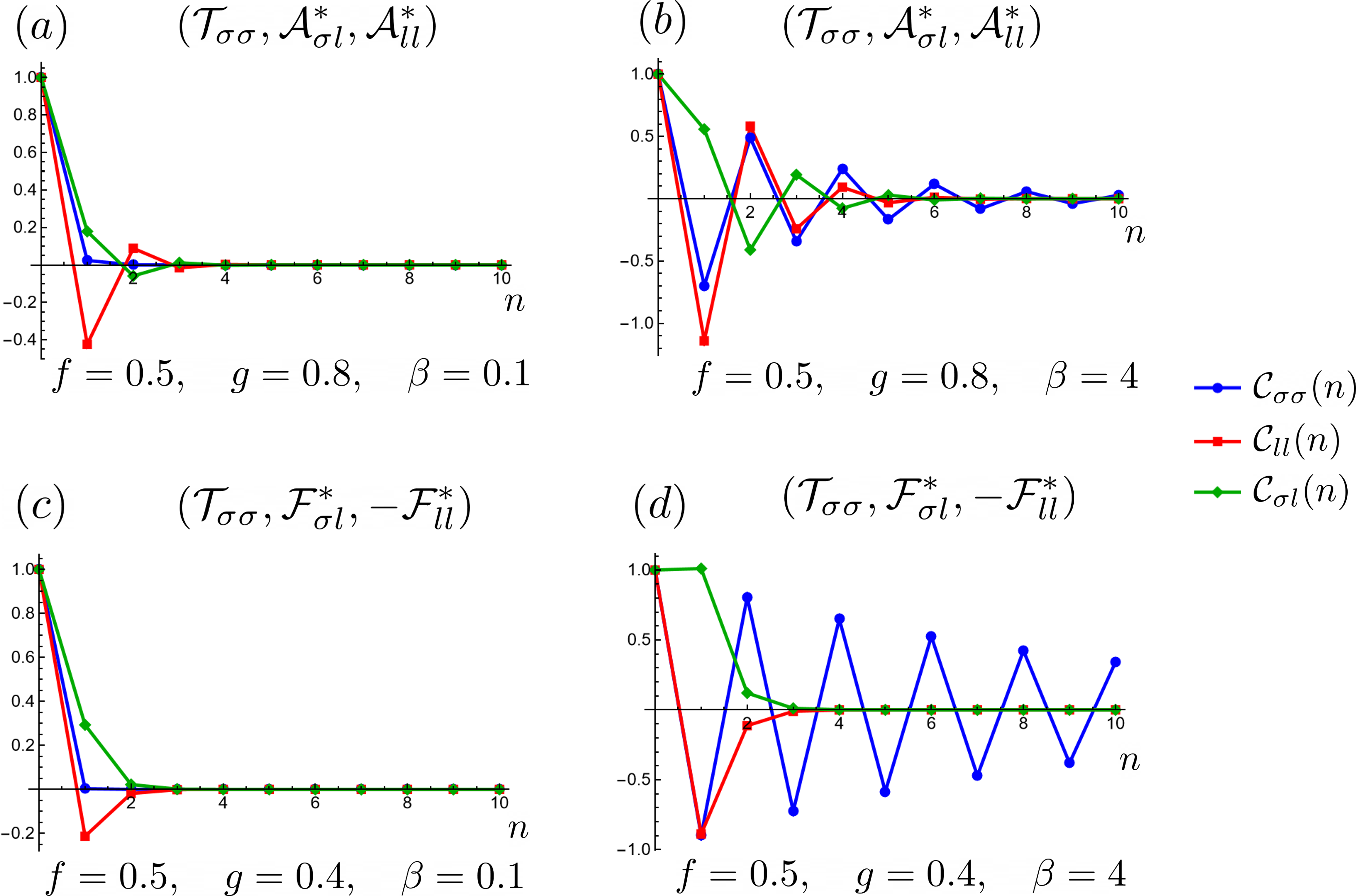}
		\captionof{figure}{\justifying \label{fig:CorrsCentre} Typical behaviour for the correlation functions in the $(\mathcal{T}_{\sigma \sigma},\mathcal{A}^*_{\sigma l},\mathcal{A}^*_{l l})$, top row, and $(\mathcal{T}_{\sigma \sigma},\mathcal{F}^*_{\sigma l},-\mathcal{F}^*_{l l})$, bottom row, of Fig.~\ref{fig:AllOrd}. Low density behaviour is shown on the left (panels (a) and (c)) while high densities are shown on the right (panels (b) and (d)). In both the top and bottom rows the orientational correlations, shown in blue, are seen to transition from ferromagnetic to antiferromagnetic with increasing density and the correlation length increases. In the upper row, the layer-layer (red) and layer-orientation (green) correlations are antiferromagnetic (i.e. consistently switching signs) in character, after the nearest neighbours. In the lower row, they are ferromagnetic (i.e. have a consistent sign), but $C_{ll}$ is negative. In each case, while it increases initially, the degree of this order becomes small at the highest packing fractions.}
	\end{figure}

In summary, the central region of Fig.~\ref{fig:OrrOrd}, is divided into two parts by the behaviour of $\mathcal{C}_{ll}$ and $\mathcal{C}_{\sigma l}$. When $g<1-f$, they both have consistent, but opposite, signs; $\mathcal{C}_{\sigma l}$ is positive and $\mathcal{C}_{ll}$ is negative. For this reason here we apply the labels $-\mathcal{F}^{*}_{ll}$ and $\mathcal{F}^*_{\sigma l}$, with the stars indicating that the order weakens at high packing fractions and the negative sign shows the layering order is anti-correlated. For larger layer spacings, with $g>1-f$, both correlations are anti-ferromagnetic in character and are small at high densities, leading to our labels, $\mathcal{A}^{*}_{ll}$ and $\mathcal{A}^*_{\sigma l}$. In Fig.~\ref{fig:CorrsCentre} we show typical behaviours for the correlation functions in each of the two sub-regions at both high and low packing fractions.

\subsubsection{\texorpdfstring{$\mathcal{A}_{\sigma\sigma}$: $g < 2f-1$}{A}}
\begin{figure}\includegraphics[width=8.5cm]{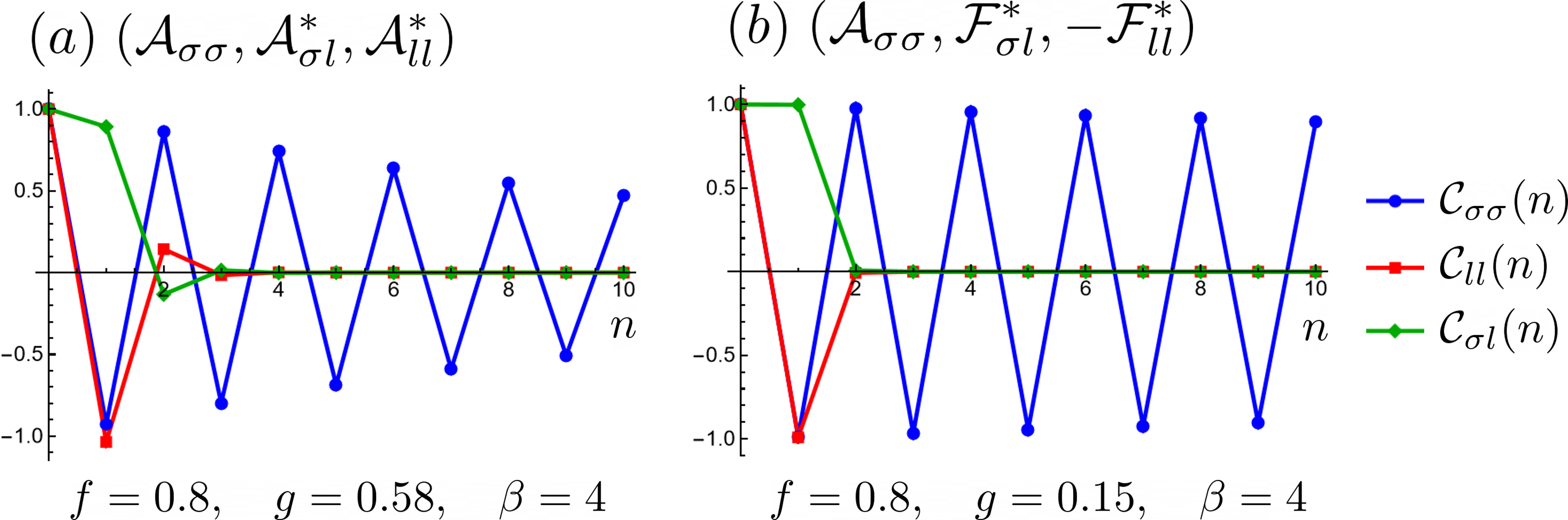}
		\captionof{figure}{\justifying \label{fig:CorrsRight} Typical behaviour of the correlation functions in the $(\mathcal{A}_{\sigma \sigma},\mathcal{A}^*_{\sigma l},\mathcal{A}^*_{l l})$, panel (a), and $(\mathcal{A}_{\sigma \sigma},\mathcal{F}^*_{\sigma l},-\mathcal{F}^*_{l l})$, panel (b), regions of Fig.~\ref{fig:AllOrd} at high density. In both cases the orientational correlations (shown in blue) are antiferromagnetic and the degree of ordering is strong. In panel (a) the layer-layer (red) and layer orientation (green) correlations are antiferromagnetic in character after the nearest neighbours, while in panel (b) they are ferromagnetic but $\mathcal{C}_{ll}$ is negative. In each case, these correlations have vanishing correlation length as packing fraction is increased.}
	\end{figure}

Finally, let us address the bottom right corner of Fig.~\ref{fig:OrrOrd} where $g<1-2f$. Here $K_2 = 1$ also, so once again we use (\ref{eq:CK2eq1}) for the correlations and our discussion is the same as above leading to the same labels;  $-\mathcal{F}^{*}_{ll}$ and $\mathcal{F}^*_{\sigma l}$ for $g<1-f$, and $\mathcal{A}^{*}_{ll}$ and $\mathcal{A}^*_{\sigma l}$ for $g>1-f$. The typical behaviours of the correlation functions in this region are shown in Fig.~\ref{fig:CorrsRight}.
    
\section{Physical Discussion of Ordering}
\label{sec:PhysDisc}
\begin{figure}\includegraphics[width=8.5cm]{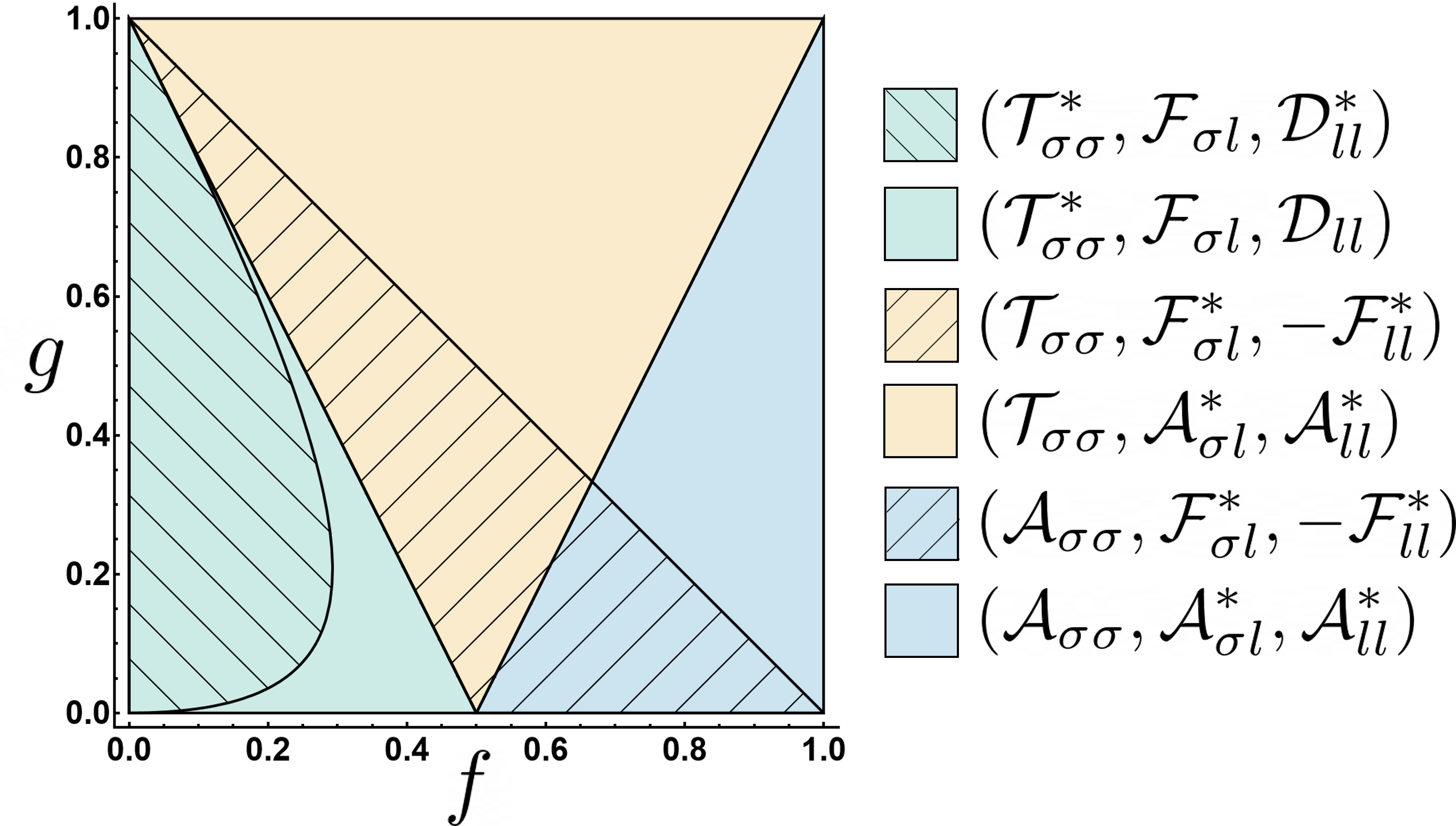}
	\captionof{figure}{\justifying \label{fig:AllOrd} The behaviour of all three correlation functions across the whole range of layer spacings, $g$, and bite sizes, $f$. The key on the right shows how the colours of the regions relate to the correlations. The symbols for the orienational correlations are the same as in Fig.~\ref{fig:OrrOrd}. The symbols $\mathcal{F}$ and $\mathcal{A}$ indicate ferromagnetic and antiferromagnetic ordering respectively. If a star is added it means that the correlation length of the order vanishes at high density. When the ferromagnetic correlations are in fact \textit{anti}-correlations, we add a negative sign. Where there is only disorder we use the symbol $\mathcal{D}$, with a star added if the correlation function noticeably changes sign.}
\end{figure}

The completed ordering diagram, adorned with all its labels showing the behaviour of all three correlations functions is given in Fig.~\ref{fig:AllOrd}. In previous sections we have outlined how this can be obtained from exact calculations. It is worth, however, trying to understand the situation physically and how this diagram can be explained without delving deeply into difficult dense derivations. That is the goal of this section, which we also divide according to the regions in Fig.~\ref{fig:OrrOrd}.

It is nigh-on impossible to explain the subtle results of our preceding analysis without repeating it. However, it remains valuable to provide some heuristic reasoning based on the lengths excluded by neighbouring particles (see Tables~\ref{tab:W0} \&~\ref{tab:Wh}) for why particular correlations are favoured in each region. 
\subsubsection{\texorpdfstring{$\mathcal{T}^{*}_{\sigma\sigma}$}{T*}: $g< 1- 2f$}
In the bottom left hand part of the $f$-$g$ plane, the particles show weak antiferromagnetic orientational ordering and almost no layering order but strong ferromagnetic layer-orientation correlations. Given that we expect there to be no layering or orientational order, let us suppose that all configurations happen with equal probability. This lets us compute a ``simple minded'' average of the excluded length between neighbors in this region, $\overline{W} = (2(1+f)-g)/4$. Compare this to the average excluded length when we enforce ferromagnetic layer-orientation correlations (i.e. $\sigma_i l_i = \sigma_{i+1}l_{i+1}$), which comes out to be $\overline{W}_{\sigma l}= f < \overline{W}$. The fact that this is \textit{smaller} than the average over all configurations shows that there is an entropic driving force for this kind of ordering; the particles are afforded more space when they correlate their layers with their orientations.  

Having demonstrated that the particles have this preference, let us investigate why they are still indifferent towards orientational and layering order in this region. We again make the restriction that $\sigma_i l_i = \sigma_{i+1} l_{i+1}$ and then compute the average excluded length when the particles have the same orientation, $\overline{W}_{\sigma l}(\text{same} \ \sigma ) = f$. This turns out to be \textit{precisely the same} as that when their orientations are different $\overline{W}_{\sigma l}(\text{diff.} \ \sigma) = f$. This shows that at the highest densities, when the orientation-layer coupling is strongest, there is no incentive for the particles to order orientationally. This points to why the anti-ferromagnetic transition for the orientational correlation functions is weakened in this region, with its correlation length approaching zero at high densities. It follows from the strong coupling between $\sigma$ and $l$, that the particles do not have any layering order either, just as we found for $\mathcal{C}_{ll}$.

\subsubsection{\texorpdfstring{$\mathcal{T}_{\sigma\sigma}$: $1-2f<g> 2f-1$}{T}}
The central region of Fig.~\ref{fig:AllOrd} harder to understand on the basis of the excluded lengths alone. Here we saw that the particles have strong anti-ferromagnetic orientational correlations at high densities while their layering order and layer-orientation correlations are weak. 

In this region, the simple minded average excluded length between a pair of particles, taken over all configurations, is $\overline{W} = (2f - 3g + 5)/8$. As we are interested in understanding antiferromagnetic orientational order, we restrict $\sigma_i \neq \sigma_{i+1}$ to find $\overline{W}(\text{diff.} \sigma) = (3-g)/4$. The difference between these two in this region is \textit{positive} $\overline{W}(\text{diff.} \sigma)-\overline{W}= (1-2f + g)/8 >0$. Indicating that, in the absence of any layering order, there is no driving force for antiferromagnetic orientational order. 

This picture can change if we take into account the layering order. This can be done by allowing there to be a small preference for the particles to be on opposite layers. Call this $p < 1/2$, so that the probability of the pair being on different layers is $1/2 + p$. Then, if this preference is large enough, ${1> 2p > (1-2f +g)/(3-2f - g)}$, our modified simple-minded averaging shows us, ${\overline{W}_p(\text{diff.} \sigma) < \overline{W}_p}$. As a reference point, take ${f=1/2=g}$. In this case, we find that $p>1/6$ would be enough to favour antiferromagnetic orientational order. 

This indicates that some layering order, in the form of a tendency for neighbouring particles to be on opposite layers, is necessary to precipitate the anti-ferromagnetic orientational order. Our exact results show that there is indeed a preference for the particles to sit on opposite layers, either due to the antiferromagnetic behaviour of $\mathcal{C}_{ll}$ or its ``negative ferromagnetic'' character.

This preference cannot be too large, however, since it does not persist to high densities. Once it has done its job in producing the orientational correlations, it must be suppressed by entropy's desire for disorder. 

 \subsubsection{\texorpdfstring{$\mathcal{A}_{\sigma\sigma}$: $g < 2f-1$}{A}}
Finally, the lower right-hand side, where the particles are almost triangles, can be understood quite easily. Recall the difference between the excluded length when a neighbouring pair of particles have different orientations and that from the average over all configurations: ${\overline{W}(\text{diff.} \sigma)-\overline{W}_0= (1-2f + g)/8}$. This is \textit{negative} in this region. Therefore, the particles can gain more freedom by assuming the opposite orientation to their neighbour which leads to the strongly antiferromagnetic behaviour of $\mathcal{C}_{\sigma \sigma}$ here. 
\section{Discussion and Conclusion}
\label{sec:conc}
In this paper, we studied the behaviour of bent-core particles on two parallel, one-dimensional layers as a function of their geometry and packing density. Each particle carried with it two Ising variables: one indicating its layer and the other its orientation. In the thermodynamic limit, we found that this model was equivalent to two coupled Ising spin-chains. This allowed us to analyse the orientation and layering order exactly, finding surprisingly rich behaviour. We introduced this model to understand how the complex phase behaviour of bent-core smectic liquid crystals can emerge in a simplified model. Now that we have understood the behaviour of the model in detail, it is worth reconnecting it to true smectics, in particular the layer-polarised $\text{SmA-P}$ states discussed in the introduction. 

First, it is convenient to translate the $\text{SmA-P}$ phases into the ordering labels used in Fig.~\ref{fig:AllOrd}. For the structure of our model to mimic the $\text{SmA-P}_{\text{F}}$ phase, all of the particles must have the same orientation, no matter the layer. This requires the correlation $\mathcal{C}_{\sigma \sigma}$ to have ferromagnetic order. From Fig.~\ref{fig:AllOrd}, we can see that this only occurs in the regions with the label $\mathcal{T}_{\sigma \sigma}$ at \textit{low density}. On the other hand, for the $\text{SmA-P}_{\text{A}}$ phase, the combination $\sigma l$ must have the same sign for each particle. This happens when the correlations $\mathcal{C}_{\sigma l}$ are ferromagnetic and is labelled in Fig.~\ref{fig:AllOrd} by $\mathcal{F}_{\sigma l}$. 

Unlike the ferromagnetic orientational order required for the $\text{SmA-P}_{\text{F}}$ phase, which occurs only at low density and has a finite correlation length, the $\text{SmA-P}_{\text{A}}$ behavior in this model appears in the green region of Fig.~\ref{fig:AllOrd} at high density, with a diverging correlation length. This observation seems qualitatively consistent with the experimental rarity of the $\text{SmA-P}_{\text{F}}$ phase and the prevalence of the $\text{SmA-P}_{\text{A}}$ phase in two-dimensional systems of bent-core particles \cite{Bisi2008,Karbowniczek2017StructureMolecules}.

Our simplified model also captures the qualitative dependence of these phases on particle geometry. For instance, two-dimensional films of bent-core particles very similar to the L-shaped bananas introduced in Sec.~\ref{sec:Bans} have been studied using density functional theory and constant-pressure Monte Carlo simulations \cite{Karbowniczek2017StructureMolecules}. These studies found that the $\text{SmA-P}_{\text{A}}$ phase emerges from the normal smectic phase as the bite size increases (or equivalent $f$ decreases) in our language. This can be seen from Fig.~\ref{fig:AllOrd} by considering the high-density behaviour. In this case, all but the green region on the left have \textit{apolar} layers, since in the yellow and blue regions the correlation functions $\mathcal{C}_{\sigma l}$ and $\mathcal{C}_{ll}$ each have vanishing correlation lengths at high density. Therefore, if we choose a layer spacing, $g$, that lies outside of the green region and start reducing $f$, the system will transition from an \textit{apolar} state to the $\text{SmA-P}_{\text{A}}$ structure.
 
It is also worth considering the other structures predicted by the ordering diagram in Fig.~\ref{fig:AllOrd}, especially the layering order. We do not find any strong layering order with diverging correlation length at high density. This is probably due to our restriction on the separation between the layers that prevents particles from escaping the interference of other on the layer above or below their own (see Fig.~\ref{fig:hmaxmin}). Nevertheless, certain ordering tendencies are found. For example, at lower densities in the \textit{un-hashed} yellow and blue regions, antiferromagnetic layering order is preferred. This is consistent with the zig-zag ordering of particles observed in cylindrical confinement \cite{Gurin2018PositionalNanopores,Gurin2015BeyondSquares}. The \textit{hashed} yellow and blue regions, on the other hand, show the particles would prefer to be two independent layers: the \textit{negative} ferromagnetic layering order shows that each particle would like to be on a different layer from each of its neighbours. This could be the signs of the transition seen for confined rods when they split into two layers, albeit impeded by our restrictions to the layer spacing. If we allowed the particles to occupy three (or more) layers then perhaps the full transition could be seen. 

This discussion illustrates that, despite its simplicity, our exactly solvable model captures qualitative features consistent with more realistic systems. However, it can, and should, be refined to incorporate additional realistic features while potentially retaining exact solvability. As alluded to above, the most natural starting point is to increase the number of layers. The equivalent system with three layers has already been solved exactly for left-right symmetric particles \cite{King2024MyDimensions}, and that is a natural starting point to improve the comparison to true smectics. 

To extend the model to a fully two-dimensional system, the number of layers would need to be increased \textit{ad infinitum}. It is not clear how such a model could be approached analytically, but a natural first step is a type of ``cell theory'' \cite{Lennard-Jones1937CriticalI,Barker1955TheLiquids,Graf1997CellSpherocylinders}. Here each particle is confined to a cell by its neighbours. The area of this cell is related to the entropy per particle and depends on the layer, orientation and geometry of the surrounding particles. This would lead to the two-dimensional generalisation of our effective Hamiltonian~\eqref{eq:HGen}. Liberated from its single-dimensional trappings, this could show a true phase transition with new and interesting features. 

Real particles also have more than two possible orientations. If they are allowed to freely rotate in the model, for example, then the Ising spins used here would be elevated to Heisenberg spins. This would lead to a something similar to the model proposed phenomenologically to understand the formation of the $\text{SmA-P}_{\text{F}}$ phase \cite{Reddy2011SpontaneousLayers}. 

Studying the response of the model system to external fields is also valuable. Not only does it provide another means of seeing true phase transitions in the system, but could offer a deeper insight into the elctro-optic or mechanical responses of these systems \cite{Kantor2009One-dimensionalNeedles}. On top of this, we have ignored the intriguing possibility to apply these simplified models to other liquid crystal phases. The quasi-one-dimensional ranks of particles in columnar phases would be ideal subjects \cite{Lopes2021PhaseCylinders}.

Simplified models, like the one studied here, open the possibility of exploring these questions in detail, without approximation. They are, however, a compromise between tractability and realism. Nonetheless, as our analysis has shown, such models can retain rich behaviour, and the insights they offer often justify the trade. 
\vspace{1mm}
\acknowledgements 
I would like to thank Prof. Randall D. Kamien for his guidance, support and encouragement while discussing this work, as well as his time reading the manuscript. This work was supported by Simons Investigator grant no. 291825 from the Simons Foundation.
\appendix
\widetext
\section{Excluded Widths Geometry}
\label{app:Geo}
In this appendix we present the geometry to obtain the excluded widths for the family of bananas introduced in section \ref{sec:Bans}. For efficiency, throughout, the relevant derivation will be outlined solely in the caption of the figure illustrating it. To begin, let us consider the range of layer spacings for these particles. For this purpose, it suffices to consider full triangles. We define the variable $g$ so that it is \textit{zero} when the layer spacing is minimum and \textit{unity} when it is maximised. In general this gives $g = (h - h_{\text{min}})/(h_{\text{max}}-h_{\text{min}})$, which reduces to (\ref{eq:gdef}) once the geometry in Figs.~\ref{fig:hminapp} \&~\ref{fig:hmaxapp}. 

\begin{minipage}{0.4\linewidth}
  \includegraphics[width=\linewidth]{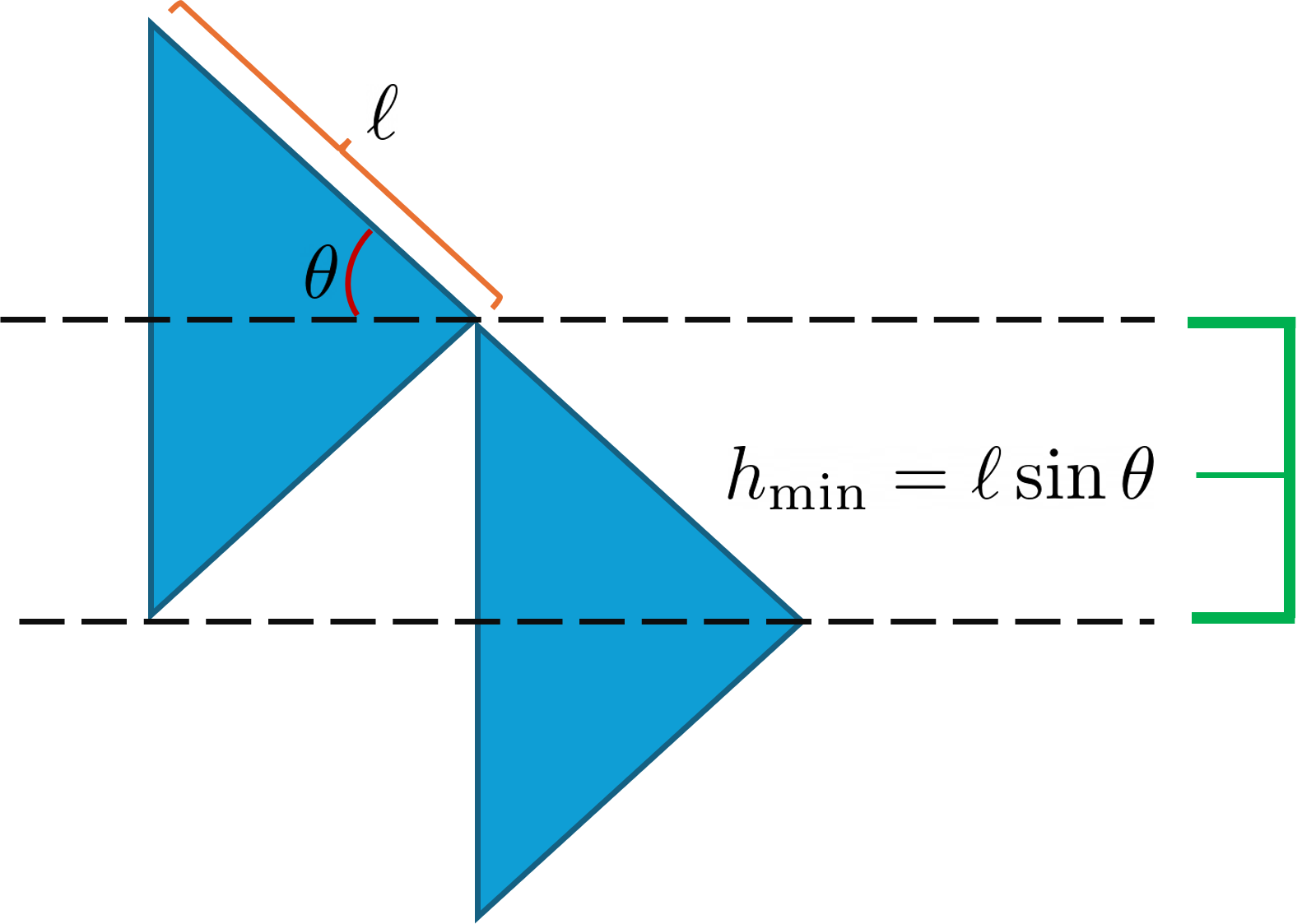}
\end{minipage}%
\hspace{0.05\linewidth}
\begin{minipage}{0.5\linewidth}
  \captionof{figure}{\justifying The smallest layer spacing, $h_{\text{min}}$, occurs when the bottom-most point of the particle on upper layer is level with the centreline with that on the lower layer. The length of the triangles' two equal sides are $\ell$, simple trigonometry yields $h_{\text{min}}= \ell \sin \theta$.}
   \label{fig:hminapp}
\end{minipage} 

\begin{minipage}{0.28\linewidth}
  \includegraphics[width=\linewidth]{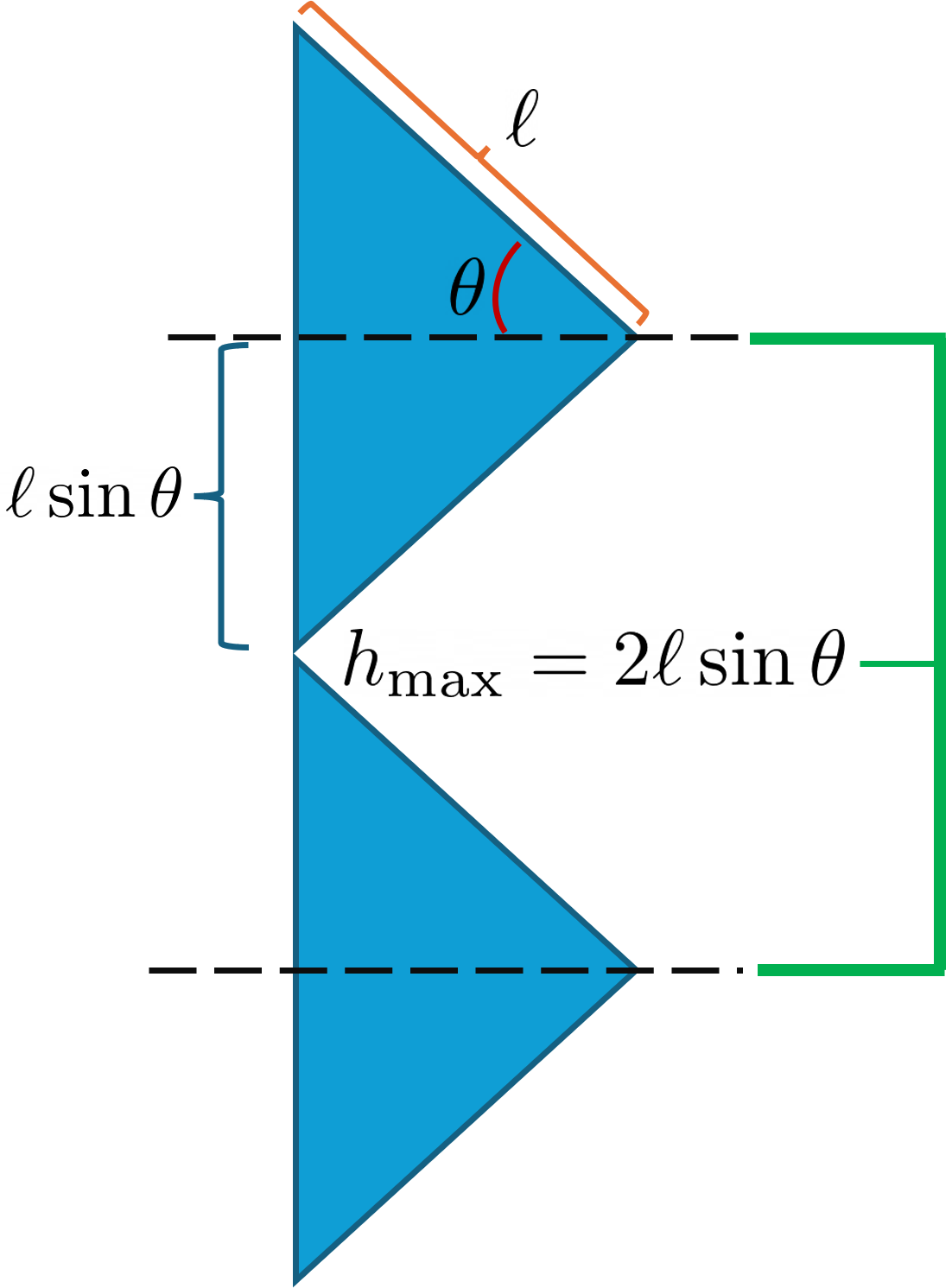}
\end{minipage}%
\hspace{0.1\linewidth}
\begin{minipage}{0.57\linewidth}
  \captionof{figure}{\justifying The largest layer spacing, $h_{\text{max}}$, when the particles only just touch at their top and bottom points. The same geometry as for the minimum spacing gives, $h_{\text{min}}= \ell \sin \theta$.}
   \label{fig:hmaxapp}
\end{minipage} 
\subsection{Same Layer}
Here we determine the excluded lengths for pairs of particles on the same layer. These are understood as the smallest separation between the leftmost points on the particles as one approaches from the left. In all that follows, the positions of the left most points are indicated by vertical red lines and that of the right-hand particle (that being approached) is marked by a tick.

\begin{minipage}{0.3\linewidth}
  \includegraphics[width=\linewidth]{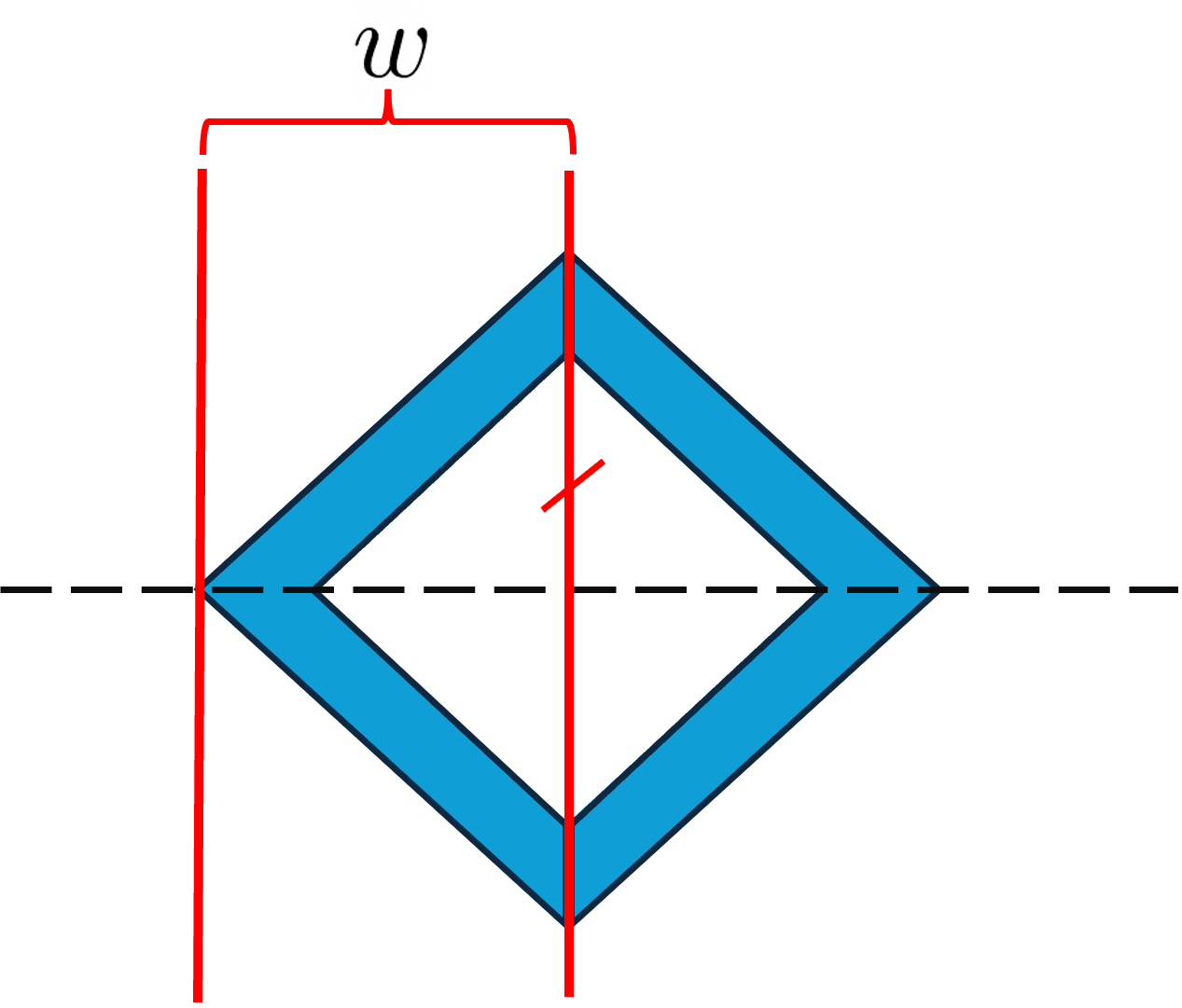}
\end{minipage}%
\hspace{0.1\linewidth}
\begin{minipage}{0.55\linewidth}
  \captionof{figure}{\justifying When $l_i = l_{i+1}$, $\sigma_{i}=-1$ and $\sigma_{i+1}=+1$, as shown here, it is clear that the excluded width is equal to that of a single particle, $w = \ell \cos \theta$.}
   \label{fig:sameldiffsapp1}
\end{minipage} \\

\begin{minipage}{0.3\linewidth}
  \includegraphics[width=\linewidth]{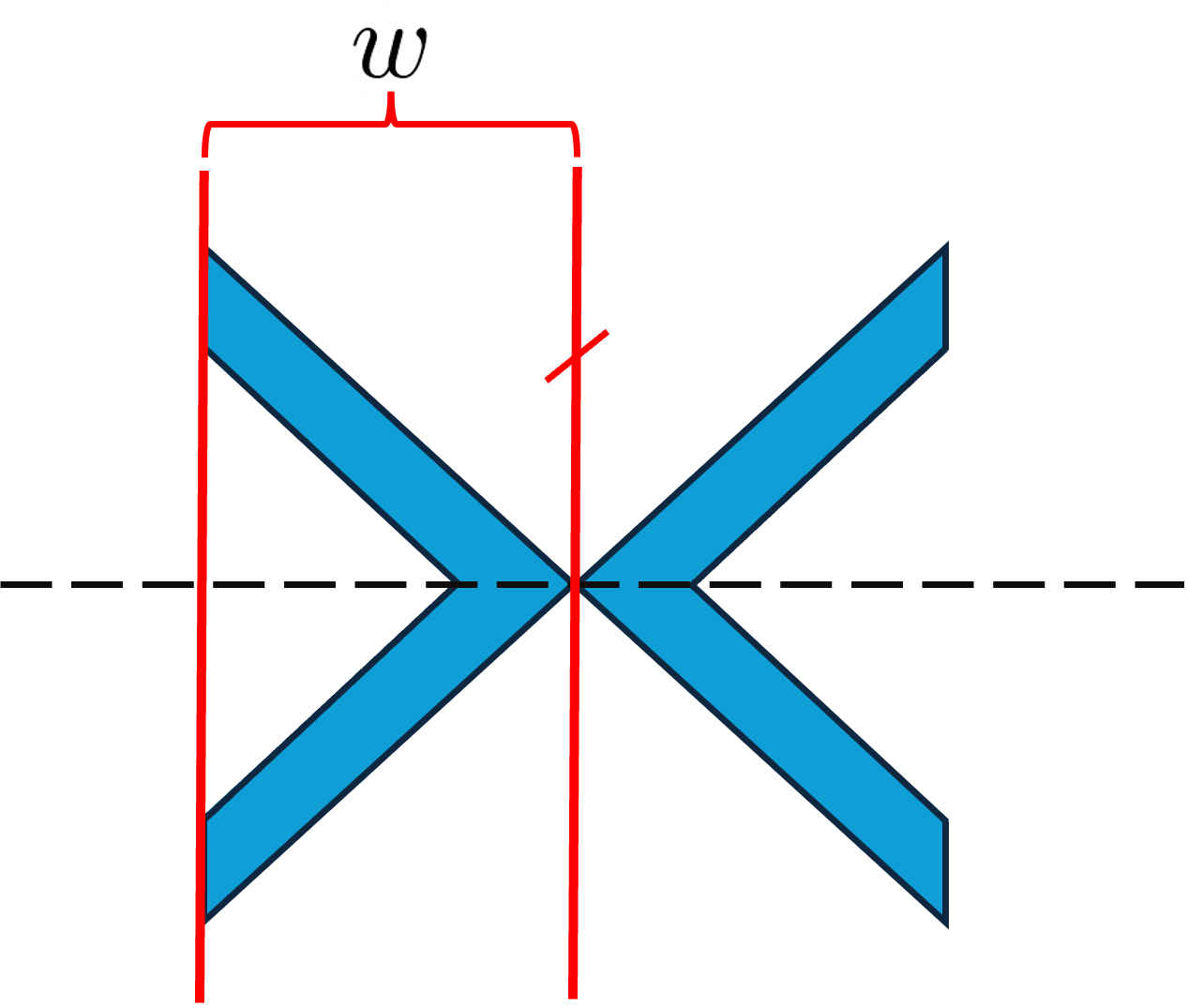}
\end{minipage}%
\hspace{0.1\linewidth}
\begin{minipage}{0.55\linewidth}
  \captionof{figure}{\justifying Here $l_i = l_{i+1}$ and $\sigma_{i}=+1$ and $\sigma_{i+1} = -1$. The excluded length here is also simply the width of the particles.}
   \label{fig:sameldiffsapp2}
\end{minipage} \\

\begin{minipage}{0.2\linewidth}
  \includegraphics[width=\linewidth]{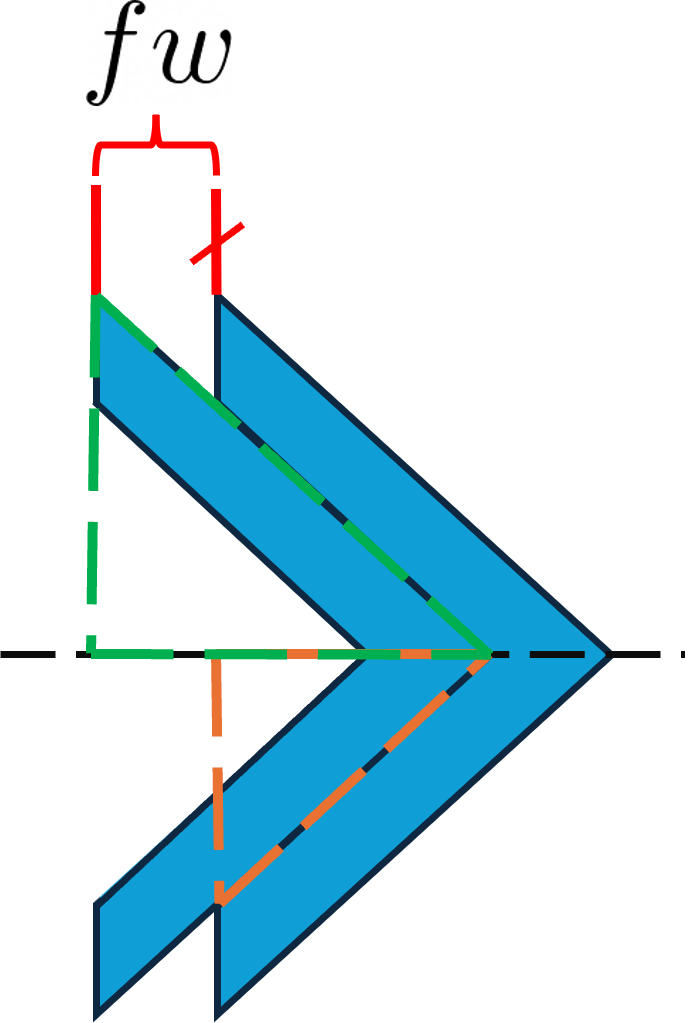}
\end{minipage}%
\hspace{0.1\linewidth}
\begin{minipage}{0.7\linewidth}
  \captionof{figure}{\justifying If the particles have a finite bite size, $f < 1$, then the excluded width between a pair on the same layer in the same orientation is \textit{smaller} than $w$. Here we show $l_i = l_{i+1}$ and $\sigma_{i}= \sigma_{i+1}= +1$, but the situation is the same for the opposite orientation. To find the excluded width, notice that the dashed green and orange triangles are similar. The ratio of their lengths is $1-f$, by definition, and the horizontal length of the green triangle is $w$. The difference between this and that of the orange triangle is the excluded length, $f w$.}
   \label{fig:samelayerapp}
\end{minipage} \\

\subsection{Different Layers}
Now we consider pairs of particles on different layers. Once again, the positions of their left-most points are indicated by vertical red lines, and the particle on the right has its line marked with a tick. Note that in one case the ticked line can appear on the left, marking the excluded length becoming \textit{negative} as discussed in Fig.~\ref{fig:K1K2Gen} of the main text. We shall only show the minimum number of cases here and the symmetries discussed in section \ref{sec:LengExc} may be used to obtain the rest. 

\begin{minipage}{0.3\linewidth}
  \includegraphics[width=\linewidth]{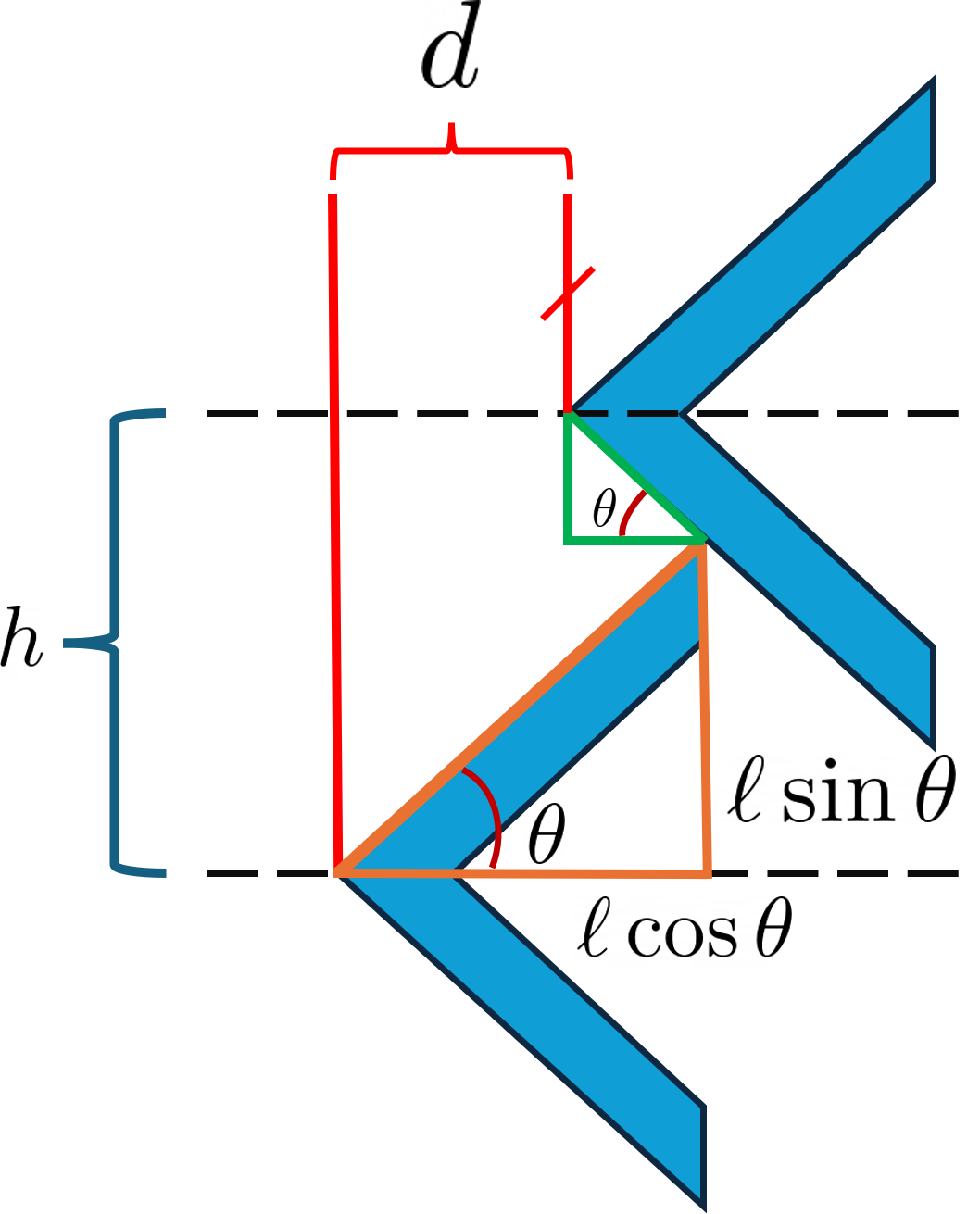}
\end{minipage}%
\hspace{0.05\linewidth}
\begin{minipage}{0.6\linewidth}
  \captionof{figure}{\justifying A good choice of similar triangles reveals the answer in the case $l_{i}=-1$, $l_{i+1}=+1$ and $\sigma_{i}=-1$, $\sigma_{i+1}=-1$. The orange and green triangles share all their angles so their lengths must be related by a constant multiple. Comparing their vertical sides one finds this multiple to be ${g = (h - \ell \sin \theta)/ \ell\sin \theta}$. The distance of closest approach is the difference between the lengths of their horizontal sides, ${d = (1 - g) w}$. Hence ${L = 1-g}$ just as in Table \ref{tab:Wh} of the main text.}
   \label{fig:difflayerapp}
\end{minipage}

\begin{minipage}{0.35\linewidth}
  \includegraphics[width=\linewidth]{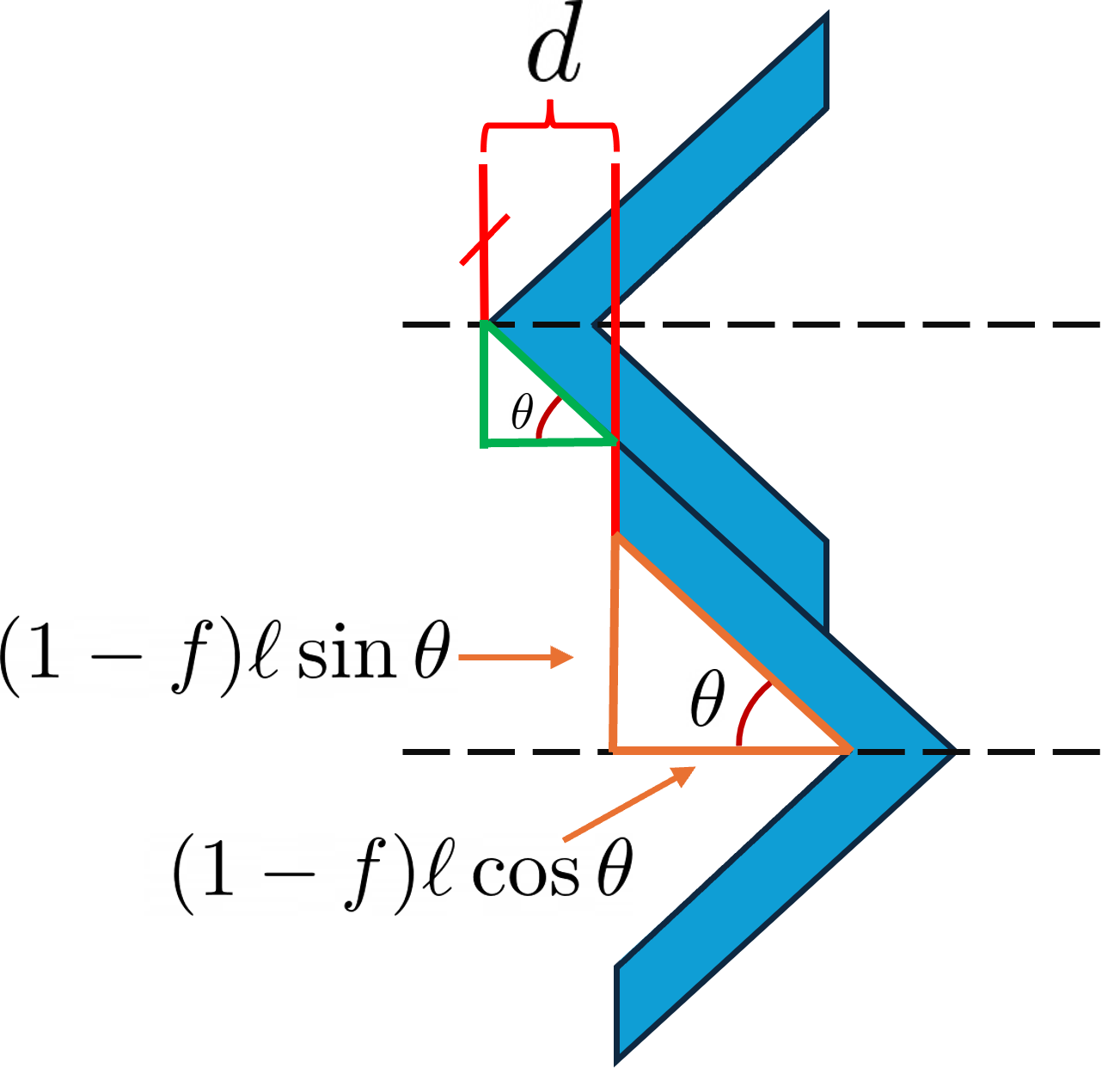}
\end{minipage}%
\hspace{0.05\linewidth}
\begin{minipage}{0.55\linewidth}
  \captionof{figure}{\justifying For $l_{i}= -1$, $l_{i+1} = +1$ and $\sigma_{i} = +1$, $\sigma_{i+1}=-1$ we can again find a useful pair of similar triangles. The vertical side of the green triangle has length $g \ell \sin \theta$, so the scale factor relating it to the orange triangle is $g/(1-f)$. Taking the difference between their horizonatal side lengths and remembering that the distance of closest approach is \textit{negative} here, we find $d= - w g$ and hence $K_1 = - g$ as in the main text. Note that this result could also be more efficiently derived by noting that $d$ must be a linear function of $g$ and $d(g=0) = 0$ and $d(g=1)=-1$. This approach will serve us well in the next case.}
   \label{fig:K1app}
\end{minipage}

\begin{center}\begin{minipage}{0.7\linewidth}
  \includegraphics[width=\linewidth]{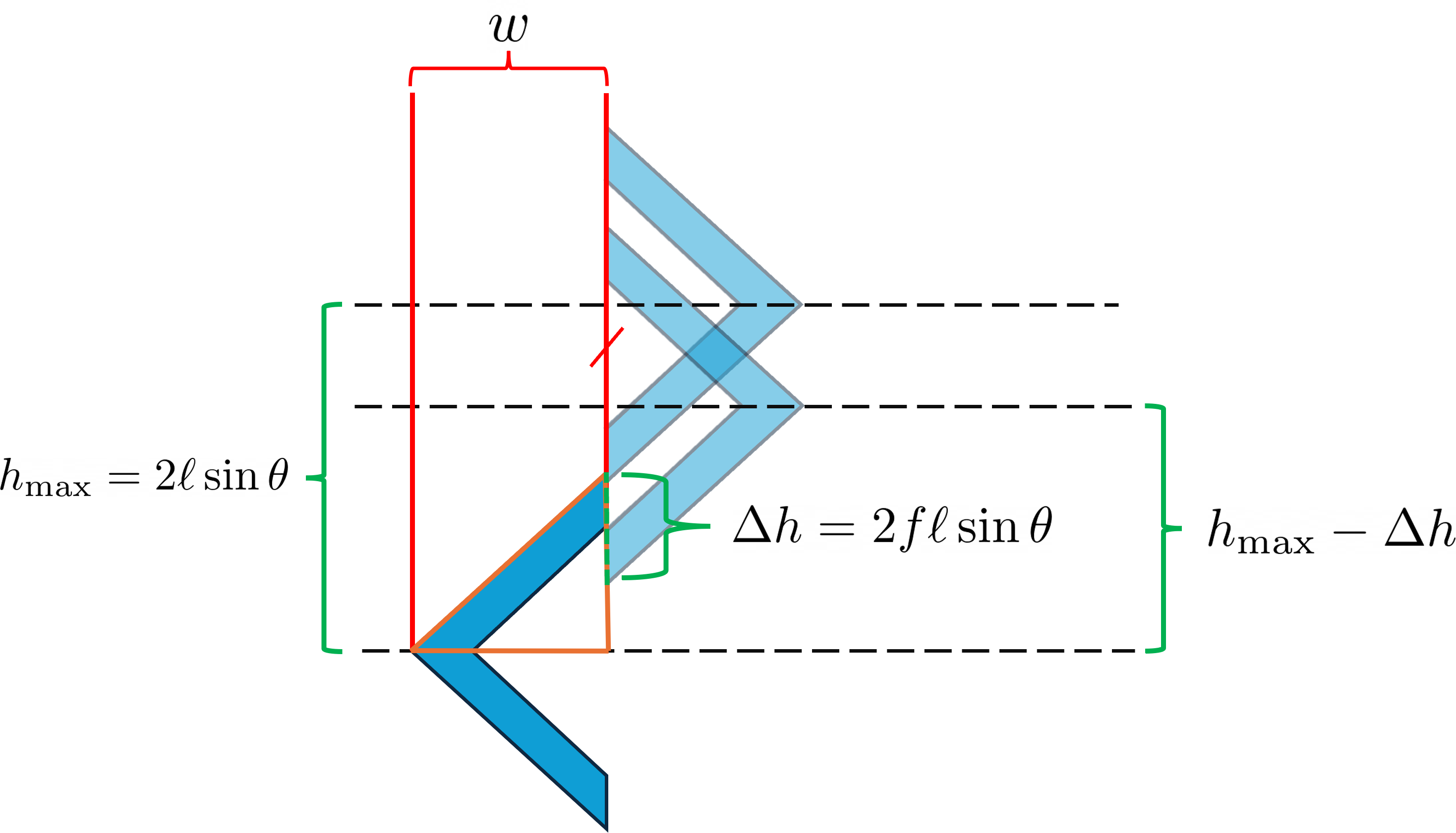}
\end{minipage}%
\captionof{figure}{\justifying \label{fig:K2app} We need to work harder for the case $l_{i}=-1$ $l_{i+1}= +1$ and $\sigma_{i}= -1$, $\sigma_{i+1}=+1$. This is because the excluded length is not a single simple function of the layer spacing. For the range of layer spacings, $\Delta h$, starting at $h_{\text{max}}$ shown above the excluded length is simply $w$. It is easy to show that $\Delta h = 2 f \ell \sin \theta$. As a function of $g$ then, we know $K_2 = 1$ if $g > 1 - 2 f$, just as in (\ref{eq:K2}) of the main text.}
\vspace{0.1cm}
\end{center}
\begin{minipage}{0.35\linewidth}
  \includegraphics[width=\linewidth]{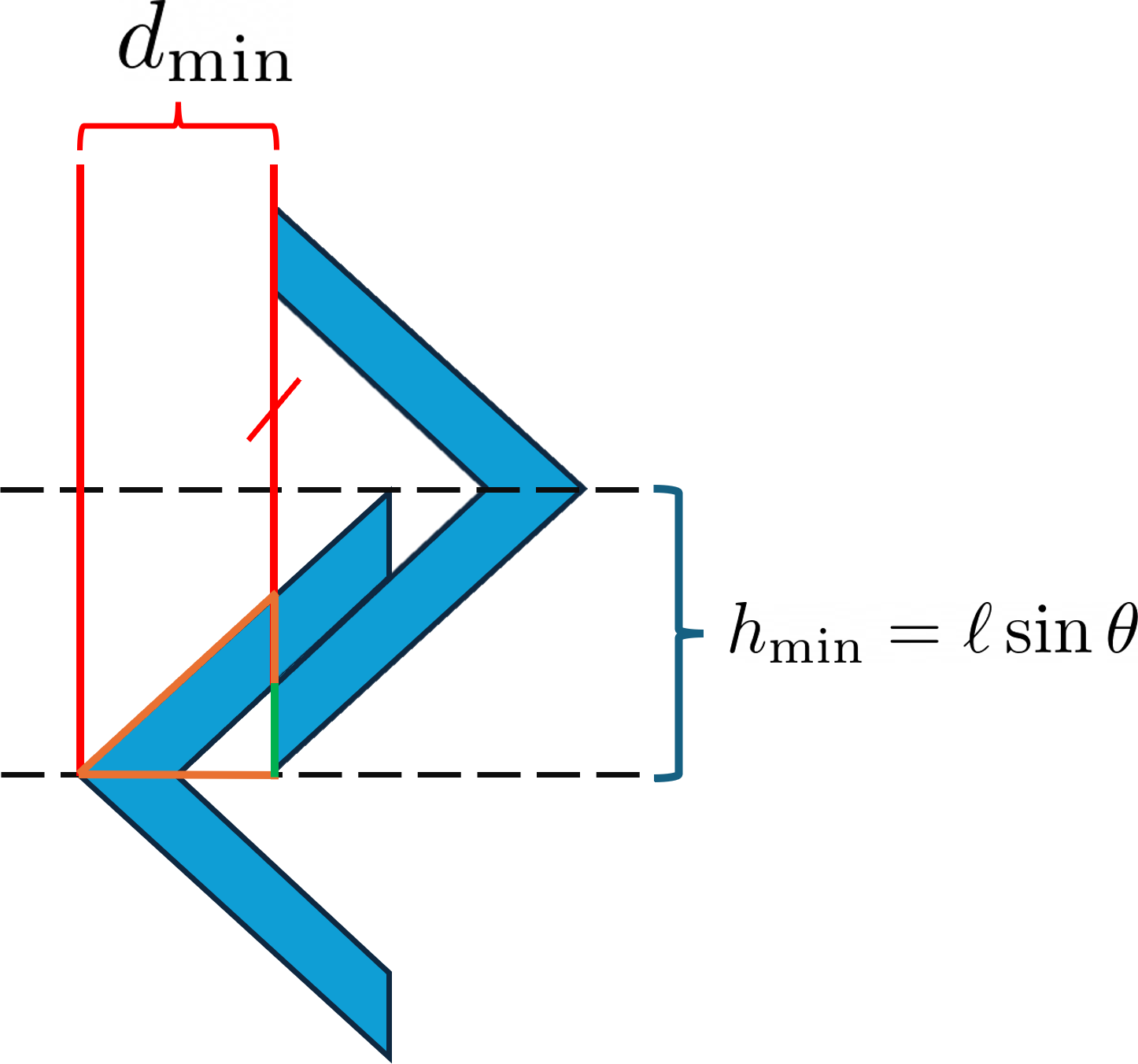}
\end{minipage}%
\hspace{0.03\linewidth}
\begin{minipage}{0.57\linewidth}
  \captionof{figure}{\justifying To deduce $K_2$ when $0<g<1-2f$ we make use of the fact that it must be a linear function of $g$, allowing us to determine it uniquely from its value at two choices of $g$. Naturally, it must be continuous, so $K_2(g =1-2f) = 1$. The second value we shall use is $K_2(0)$. As shown in this figure, there is a minimum separation between the particles at $h_{\text{min}}$ given by the length of the horizontal side of the orange triangle. We know the angles of this triangle and, from the definition of $f$, its vertical side is $2f \ell \sin\theta$. Therefore, $d_{\text{min}}= 2 f \ell \cos \theta$ and $K_2(0) = 2f$. Constructing the appropriate linear function of $g$ is straightforward and we arrive at $K_2 = g + 2 f$ if $g < 1 - 2f$, just as in the main text.}
  \label{fig:K2app1}
\end{minipage}

\section{Correlation function calculations}
\label{app:Corr}
In this appendix, we detail the calculations of the correlation functions discussed in the main text. We begin with the orientation correlations, which will provide us with our starting point for the other pair. 
\subsection{Orientational correlations\texorpdfstring{: $\mathcal{C}_{\sigma \sigma}(n)$}{ }}
\label{app:OrrCorr}
We aim to compute $\mathcal{C}_{\sigma \sigma}(n) \equiv \langle \sigma_{i} \sigma_{i+n}\rangle$, where the angle brackets denote averaging against the Boltzmann weight defined from the effective Hamiltonian (\ref{eq:HGen}). Concretely, 
\begin{equation}
    \mathcal{C}_{\sigma \sigma}(n) = \frac{\sum_{\textbf{S},\textbf{L}} \sigma_i \sigma_{i+n} e^{-\beta \mathcal{H}(\textbf{S},\textbf{L})}}{\sum_{\textbf{S},\textbf{L}}  e^{-\beta \mathcal{H}(\textbf{S},\textbf{L})}},
\end{equation}
where the denominator defines the partition function $\mathcal{Z}$. Given that the quantity whose average we seek \textit{does not} depend on the layer variables $l_i$, our first step is to sum over them. For this purpose, it suffices to consider the partition function, which we shall write as 
\begin{equation}
\label{eq:sumlbef}
    \mathcal{Z} = \sum_{\textbf{S}} e^{\sum_{i} \beta J \sigma_i \sigma_{i+1}} \sum_{\textbf{L}} e^{-\sum_i \beta \mathcal{J}(\sigma_{i},\sigma_{i+1}) l_i l_{i+1}}.
\end{equation}
The second sum over $\textbf{L}$ is exactly the partition function for a one-dimensional Ising model with a site-dependent exchange coupling $\mathcal{J}$, which can be found by changing variables to $s_i = l_i l_{i+1} = \pm 1$ so that the sum is taken trivially
\begin{equation}
\label{eq:sumlaft}
    \mathcal{Z} = 2^N \sum_{\textbf{S}} e^{\sum_{i} \beta J \sigma_i \sigma_{i+1}} \prod_{i}\cosh \left(\beta \mathcal{J}(\sigma_i,\sigma_{i+1})\right).
\end{equation}
Next, we exploit the fact that each $\sigma_{i} = \pm 1$ to write each term in the product as an exponential. Seeking a form
\begin{equation}
\label{eq:coshexp}
    \cosh \left(\beta \mathcal{J}(\sigma_i, \sigma_{i+1})\right) = \exp[a \ \sigma_{i} \sigma_{i+1} + b \ (\sigma_i - \sigma_{i+1}) + d]
\end{equation}
we find:
\begin{subequations}
    \begin{equation}
        a= \frac{1}{4}\log \frac{\cosh^2[2\beta (\alpha - L)]}{\cosh[2\beta(1-K_1)]\cosh[2\beta(1-K_2)]},
    \end{equation}
        \begin{equation}
        b= \frac{1}{4}\log \frac{\cosh[2\beta(1-K_1)]}{\cosh[2\beta(1-K_2)]},
    \end{equation}
    and
        \begin{equation}
        c= \frac{1}{4}\log \left(\cosh^2[2\beta (\alpha - L)] \cosh[2\beta(1-K_1)]\cosh[2\beta(1-K_2)]\right).
    \end{equation}
\end{subequations}
Putting this into the partition function, we have 
\begin{equation}
    \mathcal{Z} = (2 e^c)^N \sum_{\textbf{S}} \exp\left(\sum_{i} (\beta J + a) \sigma_{i} \sigma_{i+1}+  b(\sigma_{i}-\sigma_{i+1})\right).
\end{equation}
To proceed, notice that the second term in the exponential is a ``telescoping sum'', so its final result is simply $b(\sigma_N - \sigma_0)$. In the thermodynamic limit, we are free to assume periodic boundary conditions, $\sigma_N = \sigma_0$, so that this term vanishes identically. Just as discussed in section \ref{sec:OrCorr} of the main text, this results in the partition function being precisely that of an effective one-dimensional Ising model for the orientations with Hamiltonian 
\begin{equation}
\label{eq:effHapp}
    \widetilde{\mathcal{H}}(\textbf{S}) = \sum_i \widetilde{J} \sigma_i \sigma_{i+1}.
\end{equation}
just as in~\eqref{eq:Heff} of the main text. Here we have defined the effective exchange coupling 
\begin{equation}
    \widetilde{J} = J +  \frac{1}{4\beta}\log \frac{\cosh^2[2\beta (\alpha - L)]}{\cosh[2\beta(1-K_1)]\cosh[2\beta(1-K_2)]}.
\end{equation}
Having reduced things to a cornerstone, textbook problem, it is now straightforward to compute expectation values of quantities which depend solely on the orientations $\sigma_i$. 

While it is a standard result, it is worth outlining the calculation of $\mathcal{C}_{\sigma \sigma}(n)$ from here, as it serves as a simple illustration of two techniques we will use for the more complicated correlation functions. We will compute $\mathcal{C}_{\sigma \sigma}$ in three ways. Both share the same starting point, which is two exploit the property $\sigma_i^2 = 1$. This lets us write 
\begin{equation}
\label{eq:sigprod}
    \sigma_i \sigma_{i+n} = (\sigma_i \sigma_{i+1})(\sigma_{i+1} \sigma_{i+2}) \cdots (\sigma_{i+n-1} \sigma_{i+n}) = \prod_{j=i}^{n-1+i} \sigma_j \sigma_{j+1},
\end{equation}
by introducing $n-1$ factors of unity. Hence, $\mathcal{C}_{\sigma \sigma}$ can be written as the average of this $n$-fold product, and it is clear that the required sums may be taken directly, just as we did when summing over the layer variables to go from (\ref{eq:sumlbef}) to (\ref{eq:sumlaft}). Another way to go about things is less direct, but perhaps more powerful and we will find it very useful when computing more complicated correlation functions. 

To reiterate, our aim is to compute 
\begin{equation}
\label{eq:CssApp}
    \mathcal{C}_{\sigma \sigma}(n) = \frac{\sum_{\textbf{S}} (\sigma_i \sigma_{i+1})(\sigma_{i+1} \sigma_{i+2}) \cdots (\sigma_{i+n-1} \sigma_{i+n}) e^{-\beta \widetilde{\mathcal{H}}(\textbf{S})}}{\sum_{\textbf{S}} e^{-\beta \widetilde{\mathcal{H}}(\textbf{S})}}.
\end{equation}
Let us introduce a term $\mathcal{G}(\bm{\zeta}) = \beta^{-1} \sum_{i} \zeta_i \sigma_i \sigma_{i+1}$ to the Hamiltonian $\widetilde{\mathcal{H}}$ and the associated partition function $\mathcal{Z}(\bm{\zeta})$. In doing so, we have defined the vector $\bm{\zeta} = (\zeta_1, \cdots \zeta_N)$, and allowed ourselves to make the following observation
\begin{equation}
    \frac{\partial}{\partial \zeta_j} e^{-\beta(\widetilde{\mathcal{H}}(\textbf{S})+\mathcal{G}(\textbf{S}))}= \frac{\partial}{\partial \zeta_j} e^{\sum_{i}( \beta J + \zeta_i) \sigma_i \sigma_{i+1}} = \sigma_j \sigma_{j+1}  e^{-\beta(\widetilde{\mathcal{H}}(\textbf{S})+\mathcal{G}(\textbf{S}))}. 
\end{equation}
which we use to write 
\begin{equation}
\label{eq:CssPF}
    \mathcal{C}_{\sigma \sigma}(n) = \frac{1}{\mathcal{Z}}\sum_{\textbf{S}} \prod_{j=i}^{n-1+i} \lim_{\zeta_j \to 0} \frac{\partial}{\partial \zeta_j} e^{-\beta(\widetilde{\mathcal{H}}(\textbf{S})+\mathcal{G}(\textbf{S}))} = \frac{1}{\mathcal{Z}} \left(\prod_{j=i}^{n-1+i} \frac{\partial}{\partial \zeta_j}\right) \mathcal{Z}(\bm{\zeta})\bigg\lvert_{\bm{\zeta}=0}.
\end{equation}
The partition function $\mathcal{Z}(\bm{\zeta})$ is found easily by directly summing over the variables $\sigma_{i} \sigma_{i+1}$ with the result 
\begin{equation}
    \mathcal{Z}(\bm{\zeta}) = \prod_{i} \cosh (\beta \widetilde{J} + \zeta_{i}).
\end{equation}
When the required derivatives in (\ref{eq:CssPF}) are taken, $N - n$ of the terms in the above product are untouched and cancel with the same number of terms from the denominator, but the other $n$ become $\sinh(\beta \widetilde{J})$ in the limit $\bm{\zeta} \to 0$. This yeilds the well known result
\begin{equation}
\label{eq:Csigsigapp}
    \mathcal{C}_{\sigma \sigma}(n) = \frac{\sinh^n(\beta \widetilde{J})}{\cosh^n(\beta\widetilde{J})} = \left(\tanh\beta \widetilde{J}\right)^n
\end{equation}
as given in (\ref{eq:Csigsig}) of the main text. 

That was one of the traditional routes to the answer. We can also follow a less well-trodden path, which introduces another technique we will need in our armory to attack the more complex correlation functions. First, rewrite the correlation function (\ref{eq:CssApp}) in terms of a product using (\ref{eq:sigprod})
\begin{equation}
    \mathcal{C}_{\sigma \sigma}(n) = \frac{1}{\mathcal{Z}} \sum_{\textbf{S}} \left(\prod_{j=i}^{n-1+i} \sigma_{j}\sigma_{j+1} \right)e^{-\beta \widetilde{H}(\textbf{S})}. 
\end{equation}
Then, just as we did in (\ref{eq:coshexp}), we aim to write each factor $\sigma_{j} \sigma_{j+1}$ as an exponential. To do so, we must resort to complex numbers, because $\sigma_{j} \sigma_{j+1}$ can be \textit{negative}. This leaves us with an infinite number of choices, with the most natural being 
\begin{equation}
\label{eq:sscomplex}
    \sigma_{j} \sigma_{j+1} = e^{\frac{i \pi}{4} (1 - \sigma_{j}\sigma_{j+1})}.
\end{equation}
When we use this in the correlation function, we find that we are computing the ratio between two partition functions; the denominator is that of the original Ising model with exchange coupling $\beta \widetilde{J}$ at each site, while the numerator is that for an Ising model with the same exchange coupling for $N-n$ sites, but for the remaining $n$ the coupling is equal to $\beta \widetilde{J} - i\pi/4$. We have already seen how to compute these partition functions, so, remembering the factor of $e^{i \pi/4}$ from each pair $\sigma_{j}\sigma_{j+1}$ in the correlation function (see eq. (\ref{eq:sscomplex})), we get the result
\begin{equation}
    \mathcal{C}_{\sigma \sigma}(n) = \frac{\cosh^{N-n}(\beta \widetilde{J}) \ e^{in\pi/4}\cosh^{n}(\beta \widetilde{J}-i \pi/4)}{\cosh^{N}(\beta \widetilde{J})},
\end{equation}
which may be massaged into the same form as (\ref{eq:Csigsigapp}). While this method is perhaps more algebraically intensive, its advantage is that we could build the quantity whose average we sought into the Hamiltonian, allowing us to reuse the straightforward calculation of the partition function. This trick will be invaluable as we turn our attention to the layer-layer and layer-orientation correlation functions.  

\subsection{Layer-layer \& Layer-Orientation correlations\texorpdfstring{: $\mathcal{C}_{l l}(n)$ \& $\mathcal{C}_{\sigma l}(n)$}{ }}
\label{app:LLCorr}
We will need to work significantly harder to compute these correlation functions. This is because the layer variables, $l_i$, appear in the quantities to be averaged, making the step where we summed over them to find the effective Hamiltonian (\ref{eq:effHapp}) challenging. We shall follow a combination of the two paths that led to $\mathcal{C}_{\sigma \sigma}(n)$ in the previous section and begin by introducing the function 
\begin{equation}
   \mathcal{G}(\bm{\mu},\bm{\xi}) = -\beta^{-1} \sum_{i} (\mu_i l_i l_{i+1} + \xi_i \sigma_i l_i \sigma_{i+1} l_{i+1})
\end{equation}
to the Hamiltonian, as mentioned in section \ref{sec:LLCorr} of the main text. Just as before we have introduced the vectors $\bm{\mu}= (\mu_1, \cdots,\mu_N)$ and $\bm{\xi} = (\xi_1, \cdots, \xi_{N})$, and will end up with an associated partition function $\mathcal{Z}(\bm{\mu},\bm{\xi})$. Following the steps that led us to equation (\ref{eq:CssPF}) above, we see that 
\begin{equation}
\label{eq:CllCslApp}
    \mathcal{C}_{ll}(n) = \frac{1}{\mathcal{Z}}\left(\prod_{j=i}^{i+n-1}\frac{\partial}{\partial \mu_{j}}\right) \mathcal{Z}(\bm{\mu},\bm{\xi})\bigg\lvert_{\bm{\mu} = \bm{\xi}= 0}, \ \ \ \text{and} \ \ \ \mathcal{C}_{\sigma l}(n) = \frac{1}{\mathcal{Z}}\left(\prod_{j=i}^{i+n-1}\frac{\partial}{\partial \xi_{j}}\right) \mathcal{Z}(\bm{\mu},\bm{\xi})\bigg\lvert_{\bm{\mu} = \bm{\xi}= 0}.
\end{equation}
Now our task is to find $\mathcal{Z}(\bm{\mu},\bm{\xi})$. This \textit{can} be done by directly taking the sum on $l_i$, since the new variables, $\mu_i$ and $\xi_i$, simply change the defintion of the constants in $\mathcal{J}$. The result is
\begin{equation}
    \mathcal{Z}(\bm{\mu},\bm{\xi}) = 2^N \sum_{\textbf{S}} e^{\sum_{i} \beta J \sigma_i \sigma_{i+1}} \prod_{i}\cosh \left(\beta \mathcal{J}(\sigma_i,\sigma_{i+1}) - \mu_i - \xi_i \sigma_{i} \sigma_{i+1}\right).
\end{equation}
As before, we exponentiate each term in the product
\begin{equation}
    \cosh \left(\beta \mathcal{J}(\sigma_i,\sigma_{i+1}) - \mu_i - \xi_i \sigma_{i} \sigma_{i+1}\right) = \exp[a(\mu_i,\xi_i) \sigma_i \sigma_{i+1} + b(\mu_i,\xi_i)(\sigma_i - \sigma_{i+1}) + c(\mu_i,\xi_i)],
\end{equation}
and find
\begin{subequations}
    \begin{equation}
        a(\mu_i,\xi_i)= \frac{1}{4}\log \frac{\cosh^2[2\beta (\alpha - L)-\xi_i-\mu_i]}{\cosh[2\beta(1-K_1)+\xi_i-\mu_i]\cosh[2\beta(1-K_2)+\xi_i-\mu_i]},
    \end{equation}
        \begin{equation}
        b(\mu_i,\xi_i)= \frac{1}{4}\log \frac{\cosh[2\beta(1-K_1)+\xi_i-\mu_i]}{\cosh[2\beta(1-K_2)+\xi_i-\mu_i]},
    \end{equation}
    and
        \begin{equation}
        c(\mu_i,\xi_i)= \frac{1}{4}\log \left(\cosh^2[2\beta (\alpha - L)-\xi_i-\mu_i] \cosh[2\beta(1-K_1)+\xi_i-\mu_i]\cosh[2\beta(1-K_2)+\xi_i-\mu_i]\right).
    \end{equation}
\end{subequations}
This in principle allows us to define an effective Ising Hamiltonian for the orientation variables, but note that the constants which would appear therein depend on $\mu_i$ and $\xi_i$. This means that when we take the derivatives as required by (\ref{eq:CllCslApp}), we end up needing to take the average of complicated functions of the orientation variables. Concretely, we find
\begin{subequations}
\label{eq:corrsapp}
\begin{equation}
    \mathcal{C}_{ll}(n) = \left\langle\prod_{j=i}^{n-1+i}\left[a_\mu \sigma_{j} \sigma_{j+1} + b_\mu(\sigma_j - \sigma_{j+1})+ c_\mu \right] \right\rangle_{\widetilde{\mathcal{H}}}
\end{equation}
and
\begin{equation}
    \mathcal{C}_{\sigma l}(n) = \left\langle\prod_{j=i}^{n-1+i}\left[a_\xi \sigma_{j} \sigma_{j+1} + b_\xi(\sigma_j - \sigma_{j+1}) + c_\xi \right] \right\rangle_{\widetilde{\mathcal{H}}}.
\end{equation}
\end{subequations}
In these equations, we have used the shorthand notation $a_{\mu} = \partial a/\partial\mu\lvert_{\mu=\xi=0}$ and the subscript on the average indicates that it is taken against the Boltzmann weight defined by the effective Hamiltonian (\ref{eq:effHapp}). 

Before ploughing on, let us pause to note something crucial. To compute this average we need to sum over all combinations of the $\sigma_i$ variables. Let us choose one where $\sigma_{i}=-1$ and $\sigma_{i+1}=+1$. The term in $\mathcal{C}_{ll}(n)$ from this pair is $c_\mu -a_\mu - 2b_\mu$ which, after some elbow grease, comes out to be $-\tanh[2 \beta(1 - K_2)]$. Therefore, if it happens that $K_2 =1$ this term kills \textit{the entire} contribution to $\mathcal{C}_{ll}$ (and the same goes for $\mathcal{C}_{\sigma l}$). This is in complete contrast to the situation when $K_2 \neq 1$, where every combination of the orientation variables is allowed to contribute to the average.  Note that the same thing happens for $K_1$ as well. However, for the banana shaped particles we consider, $K_1 =1$ only for atypical edge cases but $K_2 =1$ for a wide range of particle shapes. This subtlety means we need a different calculation strategy for the two cases where $K_2 =1$ and $K_2 \neq 1$. We shall consider each in turn. 
\subsubsection{\texorpdfstring{: $K_2 =1$}{K2 = 1}}
Let us start by following our nose. Drawing on past experience of the Ising model, we try to cast the calculation in transfer matrix form \cite{Chaikin1995PrinciplesPhysics}. Taking $\mathcal{C}_{ll}$ for concreteness, we can write each term in the product we are averaging as a matrix
\begin{equation}
\label{eq:Qdef}
    a_\mu \sigma_{j} \sigma_{j+1} + b_\mu(\sigma_j - \sigma_{j+1})+ c_\mu  = \begin{pmatrix}
a_{\mu}+c_{\mu} & c_{\mu}-a_{\mu} +2 b_{\mu} \\
0 & a_{\mu}+c_{\mu},
\end{pmatrix} \equiv Q(\sigma_{j},\sigma_{j+1})
\end{equation}
where each entry corresponds to one of the four choices of the $\sigma_j$ and $\sigma_{j+1}$. Interpreting $e^{-\beta \widetilde{\mathcal{H}}(\sigma_i,\sigma_{i+1})}$ in the same way, and writing it as the matrix 
\begin{equation}
    T(\sigma_i,\sigma_{i+1}) =\begin{pmatrix}
e^{\beta \widetilde{J}} & e^{-\beta \widetilde{J}} \\
e^{-\beta \widetilde{J}} & e^{\beta \widetilde{J}}
\end{pmatrix},
\end{equation}
we find
\begin{equation}
\begin{split}
    &\mathcal{Z} \
    \mathcal{C}_{ll}(n) =\\ &\sum_{\sigma_{1}\cdots \sigma_N} \underbrace{T(\sigma_1,\sigma_2) \cdots T(\sigma_{i-1},\sigma_{i})}_{i \ \text{terms}}\underbrace{\left[T(\sigma_i,\sigma_{i+1}) Q(\sigma_{i},\sigma_{i+1})\right] \cdots \left[T(\sigma_{n-1},\sigma_{n}) Q(\sigma_{n-1},\sigma_{n}) \right]}_{n \ \text{terms}} \underbrace{T(\sigma_{n},\sigma_{n+1}) \cdots T(\sigma_{N-1},\sigma_{N})}_{N-n-i \ \text{terms}}
\end{split}
\end{equation}
Thus, we can write $\mathcal{C}_{ll}(n)$ as a trace
\begin{equation}
    \mathcal{Z} \ \mathcal{C}_{ll}(n) = \text{Tr}\left(T^i \cdot (T \otimes Q)^n \cdot T^{N-n-i}\right),
\end{equation}
where we have introduced the $2 \times 2$ matrix $T \otimes Q$, whose components are the products of those of $T$ and $Q$, i.e. 
\begin{equation}
    T \otimes Q = \begin{pmatrix}
e^{\beta \widetilde{J}}(a_\mu + c_\mu) & e^{-\beta \widetilde{J}}(c_\mu-a_\mu +2b_\mu) \\
0 & e^{\beta \widetilde{J}}(a_\mu + c_\mu)
\end{pmatrix}.
\end{equation}
Now, we introduce the unitary (orthogonal would also do) matrix, $U$, which diagonalises $T$;
\begin{equation}
   U = \frac{1}{\sqrt{2}}\begin{pmatrix}
-1 & 1 \\
1 & 1
\end{pmatrix},
\end{equation}
which gives
\begin{equation}
    U^{\dagger} \cdot T \cdot U = 2\begin{pmatrix}
\sinh \beta \widetilde{J} & 0 \\
0 & \cosh \beta \widetilde{J}
\end{pmatrix} \equiv D.
\end{equation}
Exploiting its unitarity (orthogonality) $U^{\dagger} U = \mathbb{I}$, we can insert it strategically into the trace, to obtain
\begin{equation}
    \mathcal{Z} \ \mathcal{C}_{ll}(n) = \text{Tr}\left(D^i \cdot  [U^{\dagger} \cdot (T\otimes Q)^n \cdot U ]\cdot D^{N-n-i}\right).
\end{equation}
Evidently the central term in the square bracket is what must be dealt with next. To do so we need the extremely useful identity
\begin{equation}
\label{eq:crucialid}
    (T \otimes Q)^n = \begin{pmatrix}
e^{n\beta \widetilde{J}}(a_\mu + c_\mu)^n & n e^{(n-2)\beta \widetilde{J}}(c_\mu-a_\mu +2b_\mu) (a_\mu + c_\mu)^{n-1}\\
0 & e^{n\beta \widetilde{J}}(a_\mu + c_\mu)^n
\end{pmatrix} \equiv e^{n \beta \widetilde{J}} (a_{\mu} + c_{\mu})^n \begin{pmatrix}
1 & -n A \\
0 & 1
\end{pmatrix},
\end{equation}
which may be proven by induction. Here we have defined $A = -e^{-2 \beta \widetilde{J}}(c_{\mu} - a_{\mu} + 2 b_{\mu})/(a_{\mu}+c_{\mu})$, which is the same as (\ref{eq:Adef}) of the main text. A few more lines of algebra and recalling $\mathcal{Z}= \cosh^N \beta \widetilde{J}$, gives us, almost, the final result
\begin{equation}
  \mathcal{C}_{ll}(n) = \frac{e^{n \beta \widetilde{J}} (a_{\mu} + c_{\mu})^n}{\cosh^n(\beta \widetilde{J})}\left[(1-n A/2)+ (1+n A/2) \tanh^{N-n}(\beta \widetilde{J}) \right].
\end{equation}
We polish this off by noting that $|\tanh x| < 1$ and $n < N$, so that, as $N \to \infty$, the final term in the square bracket vanishes and we obtain
\begin{equation}
  \mathcal{C}_{ll}(n) = \frac{e^{n \beta \widetilde{J}} (a_{\mu} + c_{\mu})^n}{\cosh^n(\beta \widetilde{J})}\left[(1-n A/2)\right],
\end{equation}
which is precisely (\ref{eq:CK2eq1ll}) of the main text. An essentially identical calculation gives us (\ref{eq:CK2eq1sl}). 
\subsubsection{\texorpdfstring{: $K_2 \neq1$}{K2 =/= 1}}
If we start to follow the same calculation when $K_2 \neq 1$ we find that the matrix $Q$ in (\ref{eq:Qdef}) has a non-zero bottom left entry. As innocuous as this sounds, this ruins the crucial identity (\ref{eq:crucialid}), making our work more arduous. So, instead, we take a different route which is somewhat more circuitous but saves us some effort. 

We start with the correlation functions as written in (\ref{eq:corrsapp}). Now that the transfer matrix approach is off the table, we turn to the trick we introduced to compute $\mathcal{C}_{\sigma \sigma}$ by exponentiating the quantity to be averaged. To that end we write
\begin{equation}
\label{eq:expapp}
    a_\mu \sigma_{j} \sigma_{j+1} + b_\mu(\sigma_j - \sigma_{j+1})+ c_\mu = e^{x_\mu \sigma_{i}\sigma_{i+1}+ y_{\mu}(\sigma_{i}-\sigma_{i+1})+z_{\mu}},
\end{equation}
to find
\begin{subequations}
\label{eq:xyzdef}
\begin{equation}
    x_{\mu} = \frac{1}{4}\log\frac{(a_{\mu}+d_{\mu})^2}{(d_{\mu}-a_{\mu}+ 2b_{\mu})(d_{\mu}-a_{\mu}- 2b_{\mu})},    
\end{equation}
\begin{equation}
    y_{\mu} = \frac{1}{4}\log\frac{(d_{\mu}-a_{\mu}+ 2b_{\mu})}{(d_{\mu}-a_{\mu}- 2b_{\mu})},    
\end{equation}
and
\begin{equation}
    z_{\mu} =\frac{1}{4}\log[(a_{\mu}+d_{\mu})^2 (d_{\mu}-a_{\mu}+ 2b_{\mu})(d_{\mu}-a_{\mu}- 2b_{\mu})].
\end{equation}
\end{subequations}
For $\mathcal{C}_{\sigma l}$, we find the same with $\mu \to \xi$. Note that this step is \textit{only} possible if no combination of $\sigma_{i}$ and $\sigma_{i+1}$ make the left hand side of (\ref{eq:expapp}) \textit{zero}, because there is no (finite) number for which $e^x=0$. This is what happened when $K_2 = 1$, and is why this route was not open to us previously. 

This makes the correlation function 
\begin{equation}
    \mathcal{C}_{ll}(n) = \left\langle \exp\left(\sum_{j=i}^{n-1+i} x_\mu \sigma_{j} \sigma_{j+1} + y_{\mu}(\sigma_{j} - \sigma_{j+1}) + z_\mu \right) \right\rangle_{\widetilde{\mathcal{H}}},
\end{equation}
which lets us pull out a factor $e^{n \beta z_{\mu}}$ and incorporate the $x_\mu$ term into the Hamiltonian by defining
\begin{equation}
\label{eq:ModEffH}
    \widetilde{\mathcal{H}}(x_{\mu}) = \beta \widetilde{\mathcal{H}} + x_{\mu}\sum_{j=i}^{n-1+i}\sigma_{j} \sigma_{j+1}.
\end{equation}
Along with this new Hamiltonian, we can define the associated partition function $\mathcal{Z}(x_\mu)$. Next, we use the telescoping property of the $y_{\mu}$ term to give $\sum_{j=i}^{n-1+i}(\sigma_{j}-\sigma_{j+1}) = \sigma_{i}-\sigma_{n}$, allowing us to write 
\begin{equation}
    \mathcal{C}_{l l}(n) =  e^{n z_{\mu}}\frac{\mathcal{Z}(x_{\mu})}{\mathcal{Z}}\Big\langle e^{y_{\mu} (\sigma_{i} - \sigma_{n+i})}\Big\rangle_{\widetilde{\mathcal{H}}(x_{\mu})}.
\end{equation}
Once again, recall that $\sigma_{i}=\pm1$ to employ the identity $e^{\pm y_{\mu}\sigma_{i}}= \cosh y_{\mu} \pm \sigma_{i}\sinh y_{\mu}$, giving
\begin{equation}
\label{eq:IntCll}
    \mathcal{C}_{l l}(n) =  e^{n z_{\mu}}\frac{\mathcal{Z}(x)}{\mathcal{Z}}\left(\cosh^2 y_{\mu} - \sinh^2y_{\mu}\big\langle \sigma_i \sigma_{i+n}\big\rangle_{\widetilde{\mathcal{H}}(x_{\mu})}  + \cosh y_{\mu} \sinh y_{\mu} \big\langle \sigma_i -\sigma_{i+n}\big\rangle_{\widetilde{\mathcal{H}}(x_{\mu})}\right).
\end{equation}
As we approach the end of our grueling work, we notice that $\big\langle \sigma_i -\sigma_{i+n}\big\rangle_{\widetilde{\mathcal{H}}_L(x_{\mu})}$ must vanish identically by symmetry. This can be seen in two ways; it is an odd function of $\sigma_{i}$ variables being averaged against an even probability distribution or, more physically, even with the modification to the effective Hamiltonian in (\ref{eq:ModEffH}), the system has a symmetry under reflection about the centre of the region between site $i$ and $i+n$. This means $\langle\sigma_i\rangle = \langle\sigma_n\rangle$, hence the average of their difference is zero. 

Now we can use standard results for the 1D Ising model given previously to read off
\begin{equation}
    \big\langle \sigma_i \sigma_{i+n}\big\rangle_{\widetilde{\mathcal{H}}(x_{\mu})} = \left[\tanh\left(\beta \widetilde{J} + x_{\mu}\right)\right]^n
\end{equation}
and 
\begin{equation}
   \frac{\mathcal{Z}(x_{\mu})}{\mathcal{Z}} = \left[\frac{\cosh(\beta \widetilde{J} + x_{\mu})}{\cosh \beta \widetilde{J}}\right]^n
\end{equation}
which, upon insertion into (\ref{eq:IntCll}) gives our final answer 
\begin{equation}
    \mathcal{C}_{l l}(n) =  \left(\frac{e^{z_{\mu}}}{\cosh \beta \widetilde{J}}\right)^n \left(\cosh^2 y_{\mu} \cosh^n (\beta \widetilde{J} +x_{\mu}) - \sinh^2y_{\mu} \sinh^n(\beta \widetilde{J}+ x_{\mu})\right).
\end{equation}
All that is left to put this in the form of equations (\ref{eq:CK2neq1}) of the main text is some tedious, albeit straightforward algebra. The $y_{\mu}$ used in this appendix is the $y$ defined in (\ref{eq:ydef}) of the main text. The hopelessly complicated forms of the decay lengths are as follows:
\begin{subequations}
\label{eq:etanuapp}
    \begin{equation}
        \eta = - \log \frac{\sinh \left(\beta J + \frac{1}{4}\log \frac{\sinh^2[2\beta (\alpha - L)]}{\sinh[2\beta(1-K_1)]\sinh[2\beta(1-K_2)]}\right)}{\cosh \left(\beta J + \frac{1}{4}\log \frac{\cosh^2[2\beta (\alpha - L)]}{\cosh[2\beta(1-K_1)]\cosh[2\beta(1-K_2)]}\right)} - \frac{1}{4}\log\left(\tanh^2[2\beta(\alpha - L)]\tanh[2\beta(1-K_1)]\tanh[2\beta(1-K_2)]\right)
    \end{equation}
        \begin{equation}
        \nu = - \log \frac{\cosh \left(\beta J + \frac{1}{4}\log \frac{\sinh^2[2\beta (\alpha - L)]}{\sinh[2\beta(1-K_1)]\sinh[2\beta(1-K_2)]}\right)}{\cosh \left(\beta J + \frac{1}{4}\log \frac{\cosh^2[2\beta (\alpha - L)]}{\cosh[2\beta(1-K_1)]\cosh[2\beta(1-K_2)]}\right)} - \frac{1}{4}\log\left(\tanh^2[2\beta(\alpha - L)]\tanh[2\beta(1-K_1)]\tanh[2\beta(1-K_2)]\right).
    \end{equation}
\end{subequations}
\section{Properties of \texorpdfstring{$\eta$ and $\nu$}{eta and nu}}
\label{app:etamu}
In this appendix we derive the asymptotic behaviour of $\eta$ and $\nu$ for large and small $\beta$. These came in handy when understanding the ordering of the particles in section \ref{sec:OrCorr}. For our purposes it suffices to determine if $\eta$ and $\nu$ are large or small in size and real or imaginary, so often we will not specify constants or ignore them all together to make the dependence on $\beta$ clear. 

First, observe that $\nu$ is always \textit{positive} and the only way $\eta$ can become imaginary is if the argument of the hyperbolic sine inside its first logarithm becomes negative. It is relatively easy to show that this cannot happen for large $\beta$. However, if we expand for small $\beta$, we find that
\begin{equation}
\label{eq:sinhconst}
    \frac{1}{4}\log \frac{\sinh^2[2\beta (\alpha - L)]}{\sinh[2\beta(1-K_1)]\sinh[2\beta(1-K_2)]} \approx \frac{1}{4}\log \left(\frac{(\alpha - L)^2}{(1-K_1)(1-K_2)}\right) + \mathcal{O}(\beta^2).
\end{equation}
This is negative if the argument of the logarithm is less than unity, and when this happens $\Im(\eta) = \pi$. This is the condition given for $\eta$ to be imaginary in section \ref{sec:Tss} of the main text. 

\subsection{\texorpdfstring{$\beta \ll 1$}{beta << 1}}
Let us consider the terms inside the first logarithms of both $\eta$ and $\mu$. The denominators are both the same, and it is not too difficult to see that they both go like $\sim \cosh k \beta \sim 1$ when $\beta \ll 1$ ($k$ is some unspecified constant here). The numerators both involve the same quantity, which we just showed behaves like a constant (\ref{eq:sinhconst}) for small $\beta$, but $\eta$ takes its hyperbolic sine and $\nu$ its hyperbolic cosine. In each case, this is a constant. Therefore, any non constant $\beta$ dependence must come from the second logarithms in (\ref{eq:etanuapp}). It follows then, since $\tanh x \sim x$ for $x \ll 1$, that 
\begin{equation}
    \eta \sim \nu \sim |\log \beta |, \ \ \ \beta \ll 1.
\end{equation}
\subsection{\texorpdfstring{$\beta \gg 1$}{beta >> 1}}
When $\beta$ is large, all of the hyperbolic tangents inside the second logarithms are close to unity, which makes the logarithm close to zero. It will turn out, then, that all of the dependence on $\beta$ comes from the first logarithms in this limit. 

To investigate this, we make use of the identity, $2\sinh(x) = e^{x}(1+e^{-2x}) \sim e^x + \mathcal{O}(e^{-2x})$ as $x\to + \infty$, and a similar one for $\cosh x$. This, along with a careful comparison of $|1-K_2|$, $|1-K_1|$ and $|\alpha - L|$, allows us to show
\begin{subequations}
\begin{equation}
    \mathbb{S}=\beta J + \frac{1}{4}\log \frac{\sinh^2[2\beta (\alpha - L)]}{\sinh[2\beta(1-K_1)]\sinh[2\beta(1-K_2)]} \sim \frac{1}{4} e^{-4 \beta (1-K_2)}\left(1-\mathcal{O}(e^{-x\beta})\right), \ \ \text{as} \ \ \beta\to \infty.
\end{equation}
\begin{equation}
\mathbb{C}=
    \beta J + \frac{1}{4}\log \frac{\cosh^2[2\beta (\alpha - L)]}{\cosh[2\beta(1-K_1)]\cosh[2\beta(1-K_2)]} \sim -\frac{1}{4} e^{-4 \beta (1-K_2)}\left(1+\mathcal{O}(e^{-x\beta})\right), \ \ \text{as} \ \ \beta\to \infty,
\end{equation}
\end{subequations}
where $x$ is some unspecified positive constant. This means both of the above are small, so Taylor expanding gives us 
\begin{subequations}
\begin{equation}
    \eta \sim -\log \frac{\sinh \mathbb{S}}{\cosh \mathbb{C}} \sim - \log \mathbb{S} \sim 4\beta (1-K_2), \ \ \text{as} \ \ \beta \to \infty,
\end{equation}
\begin{equation}
    \nu \sim -\log \frac{\cosh \mathbb{S}}{\cosh \mathbb{C}} \sim - \log \frac{1+\mathbb{S}^2/2}{1+\mathbb{C}^2/2} \sim \frac{1}{2}(\mathbb{C}^2 - \mathbb{S}^2) \sim \mathcal{O}(e^{- x \beta}), \ \ \text{as} \ \ \beta \to \infty. 
\end{equation}
\end{subequations}
While this final result is quite imprecise, it is sufficient to demonstrate that at high packing fractions, $\nu$ becomes very small, but $\eta$ becomes very large. 
\twocolumngrid
\bibliography{references}

\end{document}